\documentclass[letterpaper,twocolumn,10pt]{article}
\usepackage{usenix-2020-09}

\usepackage{tikz}
\usepackage{amsmath}
\usepackage{amssymb}    
\usepackage{verbatim}   
\usepackage{booktabs}   
\usepackage{tabularx}   
\usepackage{makecell}   
\usepackage{pifont}     
\usepackage[table]{xcolor} 
\usepackage{threeparttable}
\usepackage{xspace}    
\usepackage{algorithm}      
\usepackage{algpseudocode}
\usepackage[nameinlink]{cleveref}
\usepackage{graphicx}
\usepackage{subcaption}
\usepackage{multirow}
\usepackage{stmaryrd}
\usepackage{enumitem}

\usepackage{listings}
\usepackage{xcolor}

\usepackage{listings}
\usepackage{xcolor}

\usepackage[most]{tcolorbox}

\newcommand{\lj}[1]{\textcolor{red}{#1}}

\newcommand{\pname}{{DeRAN}\xspace}

\begin{document}

\date{}

\title{
\pname: A Neuro-Symbolic Interpretation Framework for Deep Reinforcement Learning in 5G Open RAN}

\title{
Demystifying Deep Reinforcement Learning: A Neuro-Symbolic Framework for Interpretable Open RAN Automation\vspace{-0.35in}
}
%

\author{
\vspace{1em}\\
{\rm Jie Lu}$^{\dagger}$\quad
{\rm Peihao Yan}$^{\dagger}$\quad
{\rm Pang-Ning Tan}$^{\dagger}$\quad
{\rm Y. Thomas Hou}$^{\ddagger}$\quad
{\rm Huacheng Zeng}$^{\dagger}$\\
$^{\dagger}$Michigan State University \quad
$^{\ddagger}$Virginia Tech 
}

\maketitle

\begin{abstract}

Open Radio Access Networks (O-RAN) are increasingly adopting data-driven control through Deep Reinforcement Learning (DRL) to optimize complex tasks such as network slicing and mobility management. However, the deployment of DRL in carrier-grade networks is hindered by its inherent opacity and stochastic execution, which limit operator trust, auditability, and safe deployment. Existing explainable AI (XAI) approaches primarily provide post-hoc insights and fail to produce executable, interpretable policies suitable for operational environments.
In this paper, we present \textit{\pname}, a neuro-symbolic framework that bridges the gap between DRL performance and operational transparency by distilling black-box DRL policies into human-readable symbolic representations. \pname introduces a concept-driven abstraction layer that transforms high-dimensional network telemetry into a compact set of semantically meaningful features, enabling interpretable policy learning. Building on the semantically grounded concepts, \pname synthesizes symbolic policies using deep symbolic regression (DSR) for continuous control and neurally guided differentiable logic (NUDGE) for discrete decision-making.
We implement \pname on a live 5G O-RAN testbed and evaluate it on two representative use cases. Experimental results demonstrate that DeRAN achieves 78\% and 87\% of DRL's cumulative rewards in the two use cases, while offering interpretability and auditability by design. Source code is available at \url{https://github.com/Jadejavu/DeRAN}.

\end{abstract}

\section{Introduction}
\label{Intro_1}

Open Radio Access Networks (O-RAN) are rapidly evolving toward data-driven, automated control planes \cite{polese2023understanding}. Deep Reinforcement Learning (DRL) has emerged as a powerful paradigm for optimizing complex, high-dimensional tasks such as dynamic network slicing \cite{Oranslice, yan2025near} and mobility management \cite{hmarl}. 
By learning directly from network interactions, DRL agents can achieve remarkable performance in highly dynamic environments. However, their deployment in production, carrier-grade networks faces a critical barrier: the inherent opacity of deep neural networks.
Modern DRL agents operate as \textit{black-box} controllers, mapping high-dimensional network telemetry to control actions through opaque nonlinear transformations. For network operators, deploying such unauditable systems introduces operational risks. Transparent and interpretable policies are essential to verify safety constraints and enable root-cause analysis when Quality of Service (QoS) degradation occurs. Consequently, realizing the full potential of DRL in O-RAN requires resolving the fundamental tension between performance and interpretability.

\begin{figure}[!t]
    \centering
    \includegraphics[width=\linewidth]{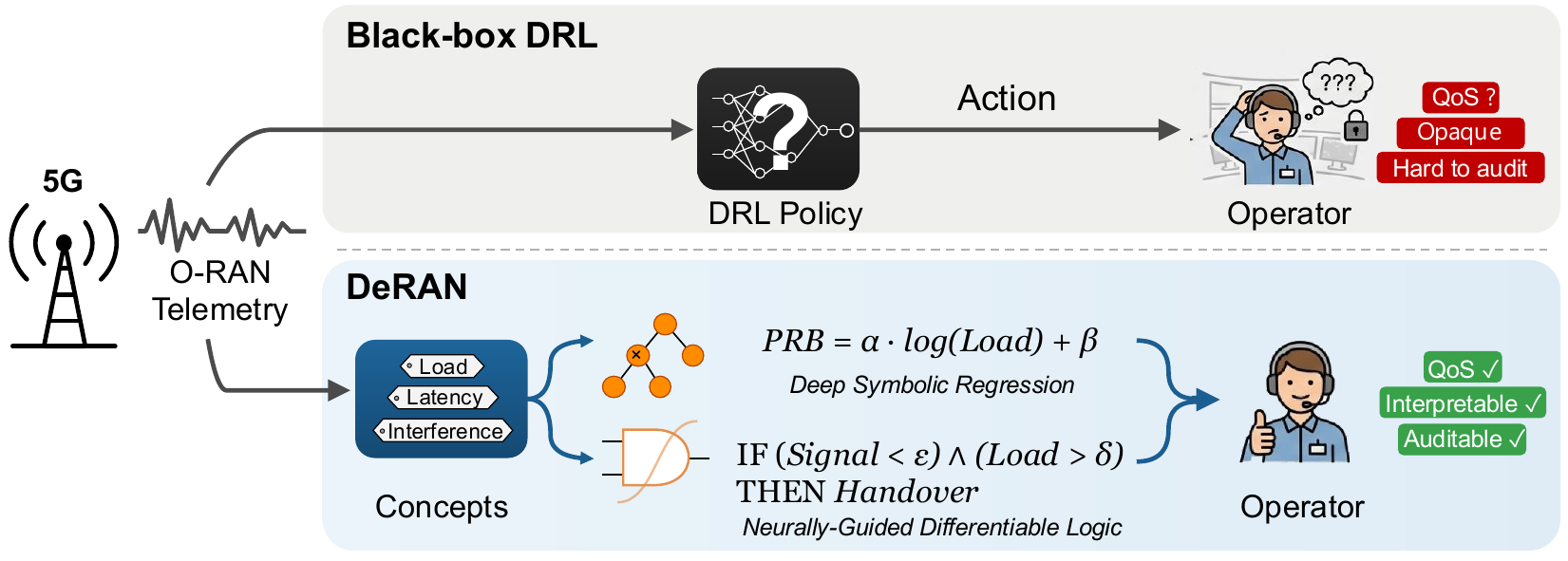} 
    \caption{\pname transforms a black-box DRL model into a white-box symbolic policy.}
    \label{fig:teaser}
\end{figure}

Existing eXplainable AI (XAI) techniques fall short of providing the level of transparency required in operational RAN environments. Post-hoc attribution methods, such as SHAP \cite{shap} and LIME \cite{lime}, quantify feature importance but do not recover the underlying decision logic of the neural network, leaving the opaque model on the critical control path. Network-specific distillation approaches, including METIS \cite{metis}, EXPLORA \cite{explora}, and ComTree \cite{comtree}, as well as symbolic auditing frameworks such as SIA \cite{sia}, attempt to extract rules directly from hundreds of raw key performance measurements (KPMs). Although these rule sets are mathematically valid, they are often large, fragmented, and operationally unintelligible, exposing a fundamental \emph{semantic gap} between low-level telemetry and the high-level predicates used by network operators. 
To date, no framework simultaneously delivers interpretable, auditable, and executable policies while preserving DRL performance and adaptability.

In this paper, we present \textit{\pname}, a neuro-symbolic interpretation framework that addresses these challenges by distilling black-box DRL policies into compact, executable, and human-interpretable representations, as shown in Fig.~\ref{fig:teaser}. Unlike conventional XAI methods that provide descriptive explanations, \pname produces intrinsically interpretable policies that can directly replace the original neural controller. The framework treats the trained DRL agent as a high-capacity teacher and extracts its decision logic into a lightweight symbolic student model, enabling deterministic and low-latency inference while preserving near-optimal performance. This transition from opaque function approximation to structured reasoning enables systematic auditing, validation, and safe deployment of AI-driven control in O-RAN systems.

A key component of \pname is its concept-driven abstraction layer, which bridges the semantic gap between raw network telemetry and human reasoning. Instead of operating on hundreds of low-level key performance measurements (KPMs), the framework discovers a compact set of high-level, semantically meaningful concepts that capture the essential network state. This abstraction is derived through a principled pipeline combining attribution analysis and ablation-driven pruning, ensuring that each concept is both interpretable and causally relevant. By projecting high-dimensional telemetry into this structured concept space, \pname significantly reduces policy complexity and enables the synthesis of concise, operator-understandable decision rules.

Building on these semantically grounded concepts, \pname synthesizes symbolic policies tailored to different action spaces. For continuous control tasks, such as resource slicing, the framework employs deep symbolic regression (DSR) \cite{dsp} to derive closed-form mathematical expressions that approximate the teacher policy. For discrete decision-making tasks, such as mobility handover, \pname leverages neurally guided differentiable logic (NUDGE) \cite{nudge} to construct compact First-Order Logic (FOL) rules that capture conditional decision structures. These symbolic policies are both interpretable and computationally efficient, making them suitable for deployment within O-RAN's near-real-time control loops.

To ensure safe and robust operation in dynamic network environments, \pname incorporates a two-stage \textit{symbolic action shielding} mechanism. The first stage performs rule-based action projection, correcting closed-form violations using domain knowledge derived from 3GPP/O-RAN specifications. The second stage introduces a safe-decision retrieval module, which replays previously verified actions when a dedicated risk estimator detects state-dependent risk.
Together, this mechanism enables \pname to rapidly respond to network dynamics while ensuring constraint compliance under uncertainty.

We implement \pname on a live multi-cell 5G NR O-RAN testbed and evaluate it on two representative DRL-based xApps: (i) resource slicing (continuous control) and (ii) UE handover (discrete control). Experimental results show that \pname achieves 78\% of the teacher DRL's cumulative reward on the resource slicing task and 87\% on the UE handover task, while delivering interpretability and auditability by design. 
More importantly, \pname also outperforms state-of-the-art XAI baselines on both control performance and policy simplicity. 
To the best of our knowledge, \pname is the first XAI framework that has been implemented and validated in an operational O-RAN environment, rather than performing offline post-hoc interpretations.

This work advances the state of the art as follows:
\begin{itemize}[leftmargin=0.15in,itemsep=0in,topsep=0in]
\item
\pname is the first neuro-symbolic framework for O-RAN that distills black-box DRL policies into \emph{intrinsically interpretable and executable} symbolic representations. 

\vspace{-0.05in}
\item
We design a learnable conceptizer that maps raw network telemetry into operator-aligned, semantically meaningful concepts. We further develop a per-dimension symbolic distillation pipeline to generate symbolic solutions for DRL policies.

\vspace{-0.05in}
\item
We implement \pname on a real-world 5G O-RAN testbed and validate its effectiveness on representative DRL xApps under realistic network dynamics.
\end{itemize}

\section{Preliminaries}
\label{Problem_3}

\subsection{O-RAN and Its DRL Policies}
\label{sec:system-model}

O-RAN introduces a disaggregated and modular architecture that decouples traditional base station functionality into distinct components, including the Radio Unit (RU), Distributed Unit (DU), Centralized Unit (CU), and the Core network. 
A key advantage of O-RAN lies in its native support for data-driven and AI-enabled control. Through open interfaces such as the E2 interface, learning-based agents can continuously collect fine-grained network telemetry in the form of KPMs, process these observations to make control decisions, and enforce actions back onto the RAN. This capability is realized through programmable control entities, namely xApps in near-RT RIC and rApps in the non-RT RIC. 
Together, xAPPs and rAPPs establish a closed-loop, AI-driven control framework that enables continuous learning and adaptation in complex and dynamic wireless environments.

\textbf{System Model.}
We consider a DRL-based xApp in O-RAN that observes the network state $\mathbf{s}_t$ and outputs a control action $\mathbf{a}_t$ at time step $t$. 
Leveraging the open and standardized interfaces of O-RAN, the xApp can collect KPMs from all entities in the network, forming a global view of the system state. Let $\mathcal{G}$ denote the set of network entities, with cardinality $G = |\mathcal{G}|$, where each entity could correspond to a user equipment (UE), a cell, or a slice, depending upon a specific task. Let $\mathcal{M}$ denote the set of KPMs extracted from each entity, with $M = |\mathcal{M}|$. 
Then, the global network state is represented as
$\mathbf{s}_t = [s_{t,g,m}] \in \mathbb{R}^{G \times M}$,
where $s_{t,g,m}$ denotes the $m$-th KPM of entity $g$ at time $t$.
Table~\ref{tab:kpm} provides an example of KPMs collected from a single entity (i.e., a UE) under two representative tasks.

\begin{table}[!t]
\centering
\caption{KPM data samples serving as the network state.}\vspace{-0.15in}
\label{tab:kpm}
\scriptsize 
\setlength{\tabcolsep}{5pt}
\begin{tabularx}{\columnwidth}{@{} l l l c X @{}}
\toprule
\textbf{Task} & \textbf{State ID} & \textbf{Metric} & \textbf{Link} & \textbf{Description / Context} \\ \midrule
\multirow{5}{*}{\rotatebox[origin=c]{90}{Slicing}} 
& $s_0, s_1$ & CQI, SNR & DL/UL & Channel quality, PUSCH SNR \\
& $s_2, s_3$ & UE PRB & DL/UL & Per-UE PRB usage \\
& $s_4, s_5$ & Throughput & DL/UL & Average data rate \\
& $s_6, s_7$ & Delay & DL/UL & RLC SDU air-interface delay \\
& $s_8, s_{9-11}$ & Vol, Slice PRB & DL & Transmitted Vol, Slice PRB usage \\ \midrule

\multirow{7}{*}{\rotatebox[origin=c]{90}{Handover}} 
& $s_{12}$ & Cell Index & -- & Primary serving cell ID \\
& $s_{13-15}$ & Srv. RSRP/Q/I & DL & Serving cell power, quality, SINR \\
& $s_{16-18}$ & Nbr. RSRP/Q/I & DL & Neighbor cell power, quality, SINR \\
& $s_{19-22}$ & Thrpt, Delay & DL/UL & Avg. data rate and RLC delay \\
& $s_{23, 24}$ & CQI, SNR & DL/UL & Channel quality and PUSCH SNR \\
& $s_{25, 26}$ & HARQ NACK & DL/UL & HARQ Nack rate \\
& $s_{27-30}$ & Srv/Nbr PRB & DL/UL & PRB usage for Srv. and Nbr. cells \\ \bottomrule
\end{tabularx}
\end{table}

Let $\pi$ be a stationary decision-making policy. The objective is to find an optimal policy $\pi^*$ that maximizes the expected cumulative discounted reward:
\begin{equation}
    \max_{\pi} \; \mathbb{E}_{\tau \sim \pi} \left[ \sum_{t=0}^{T} \gamma^{\;t} r(\mathbf{s}_t, \mathbf{a}_t) \right],
\end{equation}
where $\mathbf{a}_t \in \mathcal{A}$ is the control action, $\tau$ denotes the state-action trajectory induced by $\pi$, $\gamma \in [0,1)$ is the discount factor, and $T$ is the decision horizon. 
The step-wise reward is defined as
\[
r(\mathbf{s}_t, \mathbf{a}_t) = U(\mathbf{s}_t, \mathbf{a}_t) - \beta \, \Lambda(\mathbf{s}_t, \mathbf{a}_t),
\]
where $U(\cdot)$ captures the service utility reward, $\Lambda(\cdot)$ captures operator-defined QoS violations to penalize the reward, and $\beta \geq 0$ is a coefficient.

\textbf{DRL-Based xApp.}
Deep reinforcement learning (DRL) provides a principled framework for online decision-making in the Near-RT RIC of O-RAN. The DRL policy $\pi_\theta$ is typically implemented as a deep neural network (DNN) parameterized by $\theta$. Given the network state $\mathbf{s}_t$, the policy outputs a vector of action distribution parameters $\mathbf{z}_t \in \mathbb{R}^{d_z}$, where $d_z$ denotes the number of parameters required to specify the policy distribution (e.g., concatenated logits for discrete actions or mean and variance parameters for continuous actions).
The decision-making process is defined as
\begin{equation}
    \mathbf{z}_t = \pi_\theta(\mathbf{s}_t), 
    \quad
    \mathbf{a}_t \sim p(\cdot \mid \mathbf{z}_t),
    \label{eq:drl_mapping}
\end{equation}
where $\mathbf{z}_t$ parameterizes the action distribution $p(\cdot \mid \mathbf{z}_t)$, and $\mathbf{a}_t$ is sampled accordingly.





Although DRL has demonstrated strong performance and adaptability in O-RAN, its deployment in carrier-grade networks is hindered by inherent opacity and stochastic execution, which limit operator trust, auditability, and safe deployment.
We attribute it to the following reasons.
\emph{(i) Stochastic process:} DRL policies inherently balance exploration and exploitation; however, exploration (e.g., sampling-based action generation) introduces randomness, leading to instability and unpredictability.
\emph{(ii) Decision-making opacity:} Most DRL approaches rely on DNNs, functioning as black-box controllers that map high-dimensional network telemetry to actions through opaque nonlinear transformations.
\emph{(iii) QoS constraint violation:} DRL lacks explicit mechanisms to guarantee feasibility, and its end-to-end neural decisions may violate QoS requirements or other system-level constraints.

\subsection{A Primer on Symbolic Representations}
\label{Primer_2}

Existing XAI methods can be broadly categorized into (i) post-hoc attribution techniques (e.g., SHAP \cite{shap}, LIME \cite{lime}), which estimate feature importance for individual predictions, and (ii) symbolic approaches that produce inherently interpretable, rule-based representations. While attribution methods provide local, descriptive insights, they do not expose the underlying decision logic or offer a consistent global view. In contrast, symbolic approaches reconstruct explicit, structured policies that are transparent, verifiable, and better suited for debugging, auditing, and deployment in safety-critical systems such as O-RAN.

At a high level, symbolic methods approximate complex black-box models with interpretable surrogates that capture their input–output behavior in a structured form. This is typically achieved through techniques such as symbolic regression, program synthesis, or differentiable logic learning, which search over human-readable expressions to fit a trained model. For DRL agents, this paradigm is particularly suitable, as policies define mappings from states to actions that can often be expressed through conditional logic or low-dimensional functional relationships. By distilling these policies into symbolic forms, one can recover the implicit decision boundaries and control rules encoded in neural networks.

In the design of \pname, we will employ the following two symbolic representations.

\begin{itemize}[leftmargin=0.15in,itemsep=0in,topsep=0in]
\item 
\textit{Mathematical Expression Trees:} For continuous control tasks (e.g., power control and resource slicing), policies are represented as algebraic expressions structured as trees, where leaf nodes correspond to input variables and internal nodes represent mathematical operators (e.g., $+$, $-$, $\times$, $\max$) \cite{zhang2019novel}. Evaluating the tree yields a deterministic continuous action.

\item 
\textit{First-Order Logic (FOL):} For discrete decisions such as handover, policies are expressed as conditional IF–THEN rules. FOL formalizes these rules using atomic predicates (e.g., $\text{CQI} < 5$) and logical operators such as AND ($\land$) and OR ($\lor$) \cite{halpern2008using}, enabling transparent, verifiable, and easily modifiable decision logic.
\end{itemize}

\section{Problem Statement}
\label{sec:ioran-glance}

\begin{table}[!t]
\centering
\caption{Concept examples of resource slicing and handover.}\vspace{-0.15in}
\label{tab:semantic_mapping}
\footnotesize 
\setlength{\tabcolsep}{6pt} 
\begin{tabularx}{\columnwidth}{@{} l l l X @{}}
\toprule
\textbf{Task} & \textbf{Concept ID} & \textbf{Physical Meaning ($\eta_k$)} & \textbf{KPM Sets ($\mathcal{M}_k$)} \\ \midrule

\multirow{4}{*}{\rotatebox[origin=c]{90}{Slicing}} 
& $c_0$ & eMBB Demand & $\{s_4, s_8\}$ \\
& $c_1$ & URLLC stress & $\{s_0, s_6\}$ \\
& $c_2$ & Slice load & $\{s_{9-11}\}$ \\
& $c_3$ & Channel quality & $\{s_0\}$ \\ \midrule

\multirow{5}{*}{\rotatebox[origin=c]{90}{Handover}} 
& $c_4$ & Srv. Signal Quality & $\{s_{13-15}, s_{23}\}$ \\
& $c_5$ & Tgt. Signal Quality & $\{s_{16-18}\}$ \\
& $c_6$ & Srv. Cell Load & $\{s_{27, 28}\}$ \\
& $c_7$ & Tgt. Cell Load & $\{s_{29, 30}\}$ \\
& $c_8$ & QoS Degradation & $\{s_{19}, s_{21, 22}, s_{25}\}$ \\ \bottomrule
\end{tabularx}
\vspace{-0.2in}
\end{table}


\pname adopts a \emph{teacher-student} structure. The teacher is the DRL policy $\pi_\theta$: it continues to explore the environment as in any standard deployment and produces the high-reward decisions we want to preserve. The student $\pi_\phi$ is a transparent policy that runs on the Near-RT RIC. The teacher streams its traces $(t, \mathbf{s}_t,\mathbf{a}_t, \mathbf{z}_t)$
into a rolling buffer $\mathcal{D}$, and the student is continuously fitted to $\mathcal{D}=\{( t, \mathbf{s}_t,\mathbf{a}_t, \mathbf{z}_t)\}$ so that its distribution $\hat{\mathbf{z}}_t$ tracks the teacher's $\mathbf{z}_t$. This paradigm lets the teacher keep its learning flexibility while the student provides the three properties the teacher cannot: \textit{spec-level auditability}, \textit{deterministic execution}, and \textit{ constraint enforcement}.


\textbf{Concept Template.} 
A \emph{concept} is an expert-defined and task-level abstraction that aggregates a subset of related KPMs into a single semantically meaningful indicator. 3GPP/O-RAN specifications (e.g., 3GPP TS~28.552 \cite{3gpp.28.552}) enumerate tens of KPMs per task, but the logic reasoning operates on only a few higher-level predicates, e.g., "\texttt{target cell is strong}" or "\texttt{target slice is overloaded}". Each derived from several KPMs and their interactions (e.g., a serving-vs-neighbor RSRP comparison). 
Denote $\mathcal{C}$ as the concept template of a given control task. 
Then, it can be formulated as a three-element tuple:
\begin{equation}
    \mathcal{C} = \{(\eta_k,\mathcal{G}_k,\mathcal{M}_k)\}_{k=1}^{K},
\end{equation}
where $K=|\mathcal{C}|$ denotes the number of concepts, determined by the protocol specifications. For each concept $k$, $\eta_k$ is the semantic meaning that operators can directly audit against, i.e., "\texttt{Target-Cell Load}",  "\texttt{Per-Slice Demand}". $\mathcal{G}_k \subseteq \mathcal{G}$ is the subset of network entities related to concept $k$ (e.g., the subset of UEs belong to the eMBB slice $k$), and $\mathcal{M}_k \subseteq \mathcal{M}$ is the subset of KPMs related to concept $k$.  

At each step $t$, \pname represents the concept-level of the network by a \emph{concept vector} $\mathbf{c}_t=[c_{t,1},\ldots,c_{t,K}]^{\top} \in [0, 1]^K$, whose $k$-th entry $c_{t,k} $ is the activation value. Once the control task and its spec basis are chosen, $\mathcal{C}$ is an
expert-committed and time-invariant \textit{contract} that does not change during deployment. In contrast, $\mathbf{c}_t$ is computed at runtime from the current state $\mathbf{s}_t$ by the conceptizer (\S\ref{sec:conceptizer}), whose parameters are learned from the buffer $\mathcal{D}$. 
\emph{Experts define each concept template and its dependencies once, while \pname learns to compute its value at runtime.}

\textbf{Distillation objective.}
Given concept template $\mathcal{C}$ and data trace buffer $\mathcal{D}$, \pname trains the symbolic student policy $\pi_\phi$ to track the teacher's pre-activation outputs by minimizing
\begin{equation}
\min_{\phi} \mathcal{L}_\text{distill}(\phi) = \frac{1}{N}\sum_{t=1}^{N} d\bigl(\mathbf{z}_t,\hat{\mathbf{z}}_t\bigr),
\label{eq:distill-obj}
\end{equation}
where $d(\cdot,\cdot)$ is a distribution divergence between the teacher and student over the current buffer window $N=|\mathcal{D}|$. The optimization is performed in two stages, i.e., a \textit{spec-grounded conceptizer} followed by per-dimension \textit{symbolic distillers}, which are detailed in \S\ref{sec:overview}.


The student symbolic policy $\pi_\phi$ is designed to satisfy the following requirements.
\textit{(i) Deterministic execution.} During real-time operation, the student generates actions via a fixed component-wise head $g(\cdot)$:
\begin{equation}
\hat{\mathbf{z}}_t = \pi_\phi(\mathbf{s}_t;\mathcal{C}),
\qquad
\hat{\mathbf{a}}_t = g(\hat{\mathbf{z}}_t),
\label{eq:sym_mapping}
\end{equation}
where $g(\cdot)$ is a deterministic function, corresponding to \texttt{argmax} over categorical logits for discrete actions and the distribution mean for continuous actions.
\textit{(ii) Spec-level auditability.} The decision process is expressed as either a closed-form expression or a set of IF–THEN rules over expert-defined concepts in $\mathcal{C}$, making each control decision traceable to underlying KPMs and specification events.
\textit{(iii) Constraint compliance.} A shielding mechanism is applied to the generated action, projecting each candidate $\hat{\mathbf{a}}_t$ onto the feasible region defined by network constraints.

\section{\pname: Design}
\label{Design_4}
\subsection{Overview}
\label{sec:overview}

\begin{figure}
    \centering
    \includegraphics[width=\linewidth]{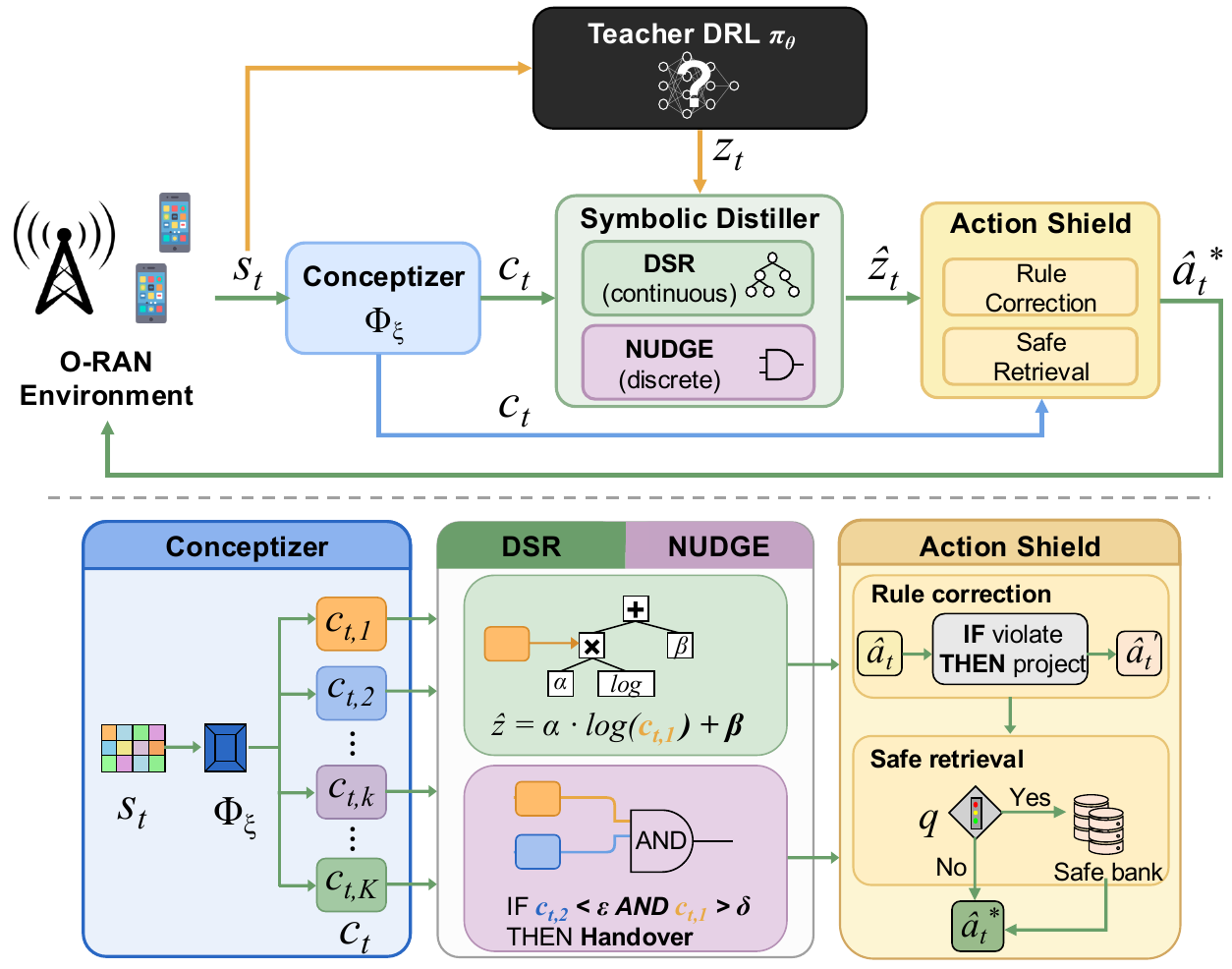}
    \caption{Overview of \pname.}
    \label{fig:ioran}
\end{figure}

Fig.~\ref{fig:ioran} illustrates the overall architecture of \pname and its interaction with the teacher DRL model at runtime. \pname consists of three sequential modules. The \emph{spec-grounded conceptizer} compresses $\mathbf{s}_t$ into a concept vector $\mathbf{c}_t$. The \emph{symbolic distiller} then maps each dimension of the teacher output $\mathbf{z}_t$ to a human-readable form (either a closed-form expression for continuous control or an IF–THEN rule set for discrete control) over $\mathbf{c}_t$. Finally, the \emph{action shielding} module projects the candidate action into the feasible region defined by operator and specification constraints before applying it to the RAN.

The design of \pname faces three key challenges. First, \textbf{spec-grounded concept inference}: operators define concepts in $\mathcal{C}$ but do not provide their numerical values, requiring the conceptizer to infer $\mathbf{c}_t$ from the teacher’s behavior using only scoped KPMs. We address this by combining architectural masking in per-concept encoders with indirect supervision via an auxiliary linear head, avoiding off-support regularization (details in \S\ref{sec:conceptizer}). Second, \textbf{multi-dimensional symbolic distillation}: jointly distilling high-dimensional actions leads to combinatorial explosion. To address this, we exploit the factorized structure of the teacher policy and distill each dimension independently, using deep symbolic regression (DSR) for continuous outputs and neurally guided differentiable logic (NUDGE) for discrete ones (\S\ref{sec:spd}). Third, \textbf{constraint enforcement}: distillation prioritizes fidelity to the teacher rather than feasibility, so the resulting actions may violate system constraints. We address this with a two-stage \emph{symbolic action shielding} mechanism, combining rule-based projection for closed-form constraints with safe-decision retrieval for state-dependent risks (\S\ref{sec:shield}).

\begin{figure}
    \centering
    \includegraphics[width=1\linewidth]{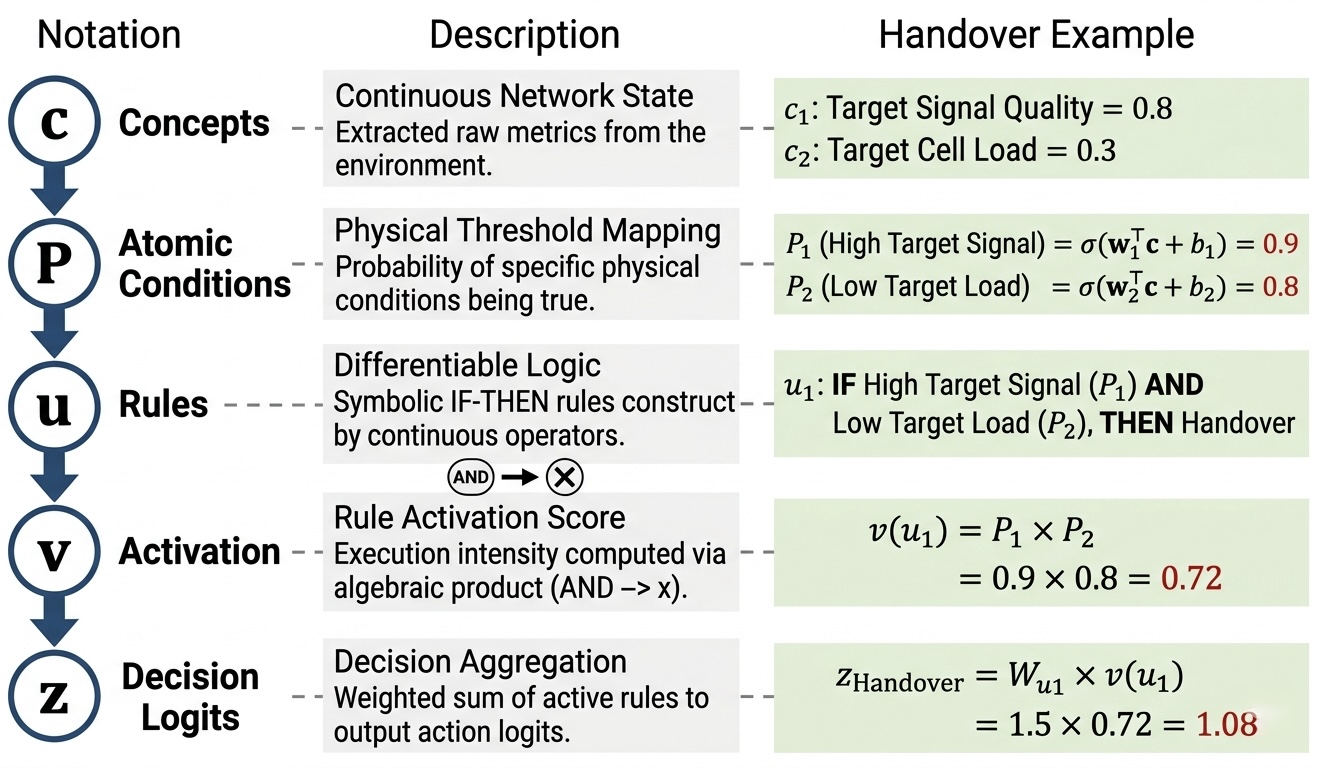}
    \caption{The logic flow of the proposed action distillation.}
    \label{fig:example_rule}
\end{figure}

Fig.~\ref{fig:example_rule} illustrates the logic flow of the proposed action distillation process, along with the key notation and a handover example.

\subsection{Spec-Grounded Conceptizer}
\label{sec:conceptizer}

A key idea of \pname is that the symbolic student operates on operator-named concepts rather than raw KPMs. 
To build semantic connection between KPMs $\mathbf{s}_t$ and concept $\mathbf{c}_t$, we define $\Phi_\xi(\cdot)$ as the conceptizer that maps each state $\mathbf{s}_t$ to a concept vector $\mathbf{c}_t = \Phi_\xi(\mathbf{s}_t) \in [0,1]^{K}$.



\textbf{Per-concept encoder and prediction head.} 
For the $k$-th concept, denote $\mathcal{M}_k$ as the set of its associated KPMs and $\mathcal{G}_k$ as the set of its associated entities.
Rather than processing the entire global state $\mathbf{s}_t$, the $k$-th concept operates only on the KPMs in $\mathcal{M}_k$ for the entities in $\mathcal{G}_k$.
Specifically, the concept value $\mathbf{c}_t = [c_{t,1}, c_{t,2}, \dots, c_{t,K}]$ is computed by
\begin{equation} 
\label{eq:concept}
    c_{t,k} = \rho_k\Bigl(\sum_{g\in\mathcal{G}_k} h_k\bigl([\mathbf{s}_{t,g}]_{\mathcal{M}_k}\bigr)\Bigr),
\end{equation}
where $[\cdot]_{\mathcal{M}_k}$ returns the subvector indexed by ${\mathcal{M}_k}$; $h_k(\cdot):\mathbb{R}^{|\mathcal{M}_k|} \to \mathbb{R}^{d_h}$ is a parameterized \textit{per-concept encoder}; and $\rho_k(\cdot): \mathbb{R}^{d_h} \to [0,1]$ is a parameterized \textit{per-concept prediction head} that maps the aggregated embedding into the concept value $c_{t,k}$. Here, $d_h$ is a shared hyperparameter across all concepts, $h_k$ and $\rho_k$ are instantiated as lightweight feedforward networks. Note that the sum in \eqref{eq:concept} is permutation-invariant following DeepSets~\cite{zaheer2017deep}. When $|\mathcal{G}_k|\!=\!1$, the sum reduces to a single term and $\rho_k$ processes the entity's embedding directly.
$\xi$ is the collection of all $\rho_k(\cdot)$ and $h_k(\cdot)$ functions in Eq.~\eqref{eq:concept}.


\textbf{Training.}
Since concept values are latent, supervision is provided indirectly from the teacher's pre-activation outputs $\mathbf{z}_t$. Specifically, we introduce an auxiliary linear head $(\mathbf{A},\mathbf{b})$ with $\mathbf{A}\in \mathbb{R}^{d_z\times K}$ that maps the concept vector back to the teacher's action-parameter space $\tilde{\mathbf{z}}_t \;=\; \mathbf{A}\,\mathbf{c}_t + \mathbf{b}$. 
The conceptizer is trained by minimizing the fidelity loss 
\begin{equation}
 \min_{\xi,\mathbf{A},\mathbf{b}} \; \mathcal{L}_\text{fid} =  \frac{1}{|\mathcal{B}|} \sum_{t\in\mathcal{B}} \bigl\|\tilde{\mathbf{z}}_t-\mathbf{z}_t\bigr\|_2^{2},
\label{eq:lfid}
\end{equation}
which requires $\mathbf{c}_t$ to preserve sufficient information to reconstruct the teacher's decisions on a batch $\mathcal{B} \subseteq \mathcal{D}$. After training, the auxiliary head $(\mathbf{A}, \mathbf{b})$ is discarded, since it serves only as a differentiable surrogate for supervising the conceptizer. The conceptizer parameters $\xi$ are then frozen, and the symbolic distillers in \S\ref{sec:spd} take $\mathbf{c}_t$ as input.

\textbf{Post-hoc auditing.}
Post-hoc auditing addresses two questions that operators can examine via Integrated Gradients (IG)~\cite{ig}: (i) whether the KPMs within $\mathcal{M}_k$ influence $c_{t,k}$ in a manner consistent with the semantic name $\eta_k$, and (ii) which KPM contributes most to the concept value at a given sample. 
Let $\mathbf{s}_\text{base}$ denote a baseline state.
Let $\mathrm{IG}_k(\mathbf{s}_t)\in\mathbb{R}^{|\mathcal{G}_k|\times|\mathcal{M}_k|}$ be a tensor-valued attribution matrix.
Its entry $(g,m)$ quantifies how strongly  KPM $[\mathbf{s}_t]_{g,m}$ 
influences the value of concept $c_{t,k}$. 
Computed over $N_\text{ig}$ interpolation steps, this attribution is given by
\begin{equation}
    [\mathrm{IG}_k(\mathbf{s}_t)]_{g,m} = \bigl([\mathbf{s}_t]_{g,m}-[\mathbf{s}_\text{base}]_{g,m}\bigr)\cdot \frac{1}{N_\text{ig}}\sum_{\ell=1}^{N_\text{ig}} \frac{\partial c_{t,k}\bigl(\mathbf{s}^{(\ell)}\bigr)}{\partial[\mathbf{s}^{(\ell)}]_{g,m}},
\end{equation}
where $g\in\mathcal{G}_k$, $m\in\mathcal{M}_k$, and $\mathbf{s}^{(\ell)} = \mathbf{s}_\text{base} + \tfrac{\ell}{N_\text{ig}}(\mathbf{s}_t - \mathbf{s}_\text{base})$ is the $\ell$-th point on the linear path between $\mathbf{s}_\text{base}$ and $\mathbf{s}_t$.

\subsection{Symbolic Distillation}
\label{sec:spd}

Based on the semantically-grounded concepts, \pname\ is then to extract the domain knowledge embedded within the black-box policy $\pi_\theta$ and synthesize an executable symbolic policy. RAN control typically involves a large multi-dimensional action space, and directly searching for a joint symbolic policy across all action dimensions simultaneously leads to a combinatorial explosion in the hypothesis space. To address this issue, we leverage the \emph{factorized} architecture of DRL 
policies, which parameterize independent marginal distributions, yielding a concatenated global parameter vector $\mathbf{z}_t$. Specifically, we decompose the distillation into parallel procedures along each action dimension. The physical correlations and joint constraints across dimensions are subsequently enforced \textit{a posteriori} by an action shielding module (\S\ref{sec:shield}). 

At each step $t$, the teacher policy $\pi_\theta$ outputs a vector $\mathbf{z}_t$, which factorizes along the action dimension $i$ as $\mathbf{z}_{t,i} \in \mathbb{R}^{d_i}$. Each $\mathbf{z}_{t,i}$ parameterizes the marginal distribution of $a_{t,i}$:
\begin{itemize}[leftmargin=1.2em,itemsep=0pt,topsep=1pt]
    \item 
    \textit{Continous Actions:} \ \ 
    When $\mathcal{A}_i \subseteq \mathbb{R}$ is continuous, we treat the teacher as a deterministic policy and distill the mean only.  Thus $d_i = 1$ and $\mathbf{z}_{t,i}$ reduces to a scalar $i$ that specifies the mean of the policy distribution;
    
    \item 
    \textit{Discrete Actions}: \ \ 
    When $\mathcal{A}_i$ is a discrete set, $d_i = |\mathcal{A}_i|$ and 
    $\mathbf{z}_{t,i}$ holds the categorical logits over $\mathcal{A}_i$.
\end{itemize}
The teacher's per-dimension action $a_{t,i}$ is then sampled from this distribution governed by $\mathbf{z}_{t,i}$.

\subsubsection{Distillation for Continuous Actions}

For continuous-action tasks such as transmit power control and resource slicing, \pname\ adapts DSR~\cite{dsp} to generate the mathematical expressions that approximate the teacher's decision function over $\mathbf{c}_t$. The core 
idea of DSR is to employ an autoregressive neural network, referred to as the \emph{expression generator}, to search the discrete space of mathematical formulas, thereby transforming equation discovery into a sequence generation problem.

\textbf{Expression Construction.} 
We instantiate the dimension-specific continuous student policy $\pi^\text{con}_i$ directly as an expression tree, whose leaf variables are the entries of the concept vector $\mathbf{c}_t$ and the internal nodes are mathematical operators. Equivalently, $\pi^\text{con}_i\!:\![0,1]^K\!\to\!\mathbb{R}$ is a deterministic scalar function that the expression generator synthesizes as a sequence of operators and variables. Evaluating the tree on the current concept vector yields the student's prediction $\hat{z}_{t,i}=\pi^\text{con}_i(\mathbf{c}_t)$.
To satisfy the timing requirements for Near-RT RIC execution, we restrict the operator set to the set below:
\begin{equation}
\mathcal{O} = \{+, \; -, \; \times, \; \div, \; \log(\cdot), \; \exp(\cdot)\}.
\end{equation}

\textbf{Data-Driven Structural Search.} 
The original DSR formulation relies on extensive trial-and-error interaction with the environment, which would bring unacceptable QoS violations in production networks. We therefore reformulate DSR into a safe and asynchronous extraction procedure that operates entirely on the streaming telemetry buffer $\mathcal{D}=\{( t, \mathbf{s}_t, \mathbf{z}_t,\mathbf{a}_t)\}$. Each candidate tree is evaluated by a behavioral fidelity reward against the teacher's target ${z}_{t,i}$:
\begin{equation}
J(\pi^\text{con}_i) = \Bigl(1+\tfrac{1}{N}\!\sum_{t=1}^{N}\bigl( \hat{z}_{t,i}-z_{t,i}\bigr)^{2}\Bigr)^{-1},
\end{equation}
where $\hat{z}_{t,i}=\pi^\text{con}_i(\mathbf{c}_t)$.
Maximizing $J(\pi^\text{con}_i)$ drives the search toward more faithful expressions to reproduce the high-performance continuous decisions of the black-box DRL policy $\pi_\theta$. To keep the generated formulas human-readable, we prune candidate formulas whose tree depth exceeds the maximum tree depth $D_\text{max}$ during search. The highest-reward expression is adopted as the interpretable initialization policy $\pi^{\text{con}}$ for the action under consideration.

\textbf{An Example.} To illustrate this procedure, consider a dynamic resource slicing task in O-RAN, where the teacher policy allocates a fractional PRB ratio to each network slice. Focusing on a slice $i$ (e.g., an eMBB slice), the teacher outputs a scalar $z_{t,i}$ denoting its PRB-ratio allocation. Suppose the conceptizer has produced two concepts: $c_{t,1}$: "\texttt{Aggregate Cell Load}" summarizes the cell-wide congestion level, and $c_{t,2}$: "\texttt{Slice Queue Backlog}" captures the aggregate queuing pressure across slices. The DSR generator combines these variables and operators to synthesize a transparent symbolic expression for action dimension $i$, for example:
$\hat{z}_{t,i} = \frac{1}{c_{t,1}}\log(1 + c_{t,2})$.
This expression admits a direct operational meaning: \textit{the allocated bandwidth share grows logarithmically with the slice's queue backlog and scales inversely with the cell-wide load}. More generally, the synthesized tree yields the deterministic continuous control output $\hat{z}_{t,i} = \pi^\text{con}_i(\mathbf{c}_t)$, making the 
teacher's decision on action dimension $i$ explicit and auditable.

\subsubsection{Distillation for Discrete Action}

For discrete actions such as beam selection or UE handover, \pname distills the teacher's marginal $\mathbf{z}_{t,i} \in \mathbb{R}^{|\mathcal{A}_i|}$ into a set of \texttt{IF-THEN} rules over the concept vector~$\mathbf{c}_t$. The challenge is that the student should be \textit{symbolic} for operators to audit, yet
\textit{differentiable} to fit the teacher's distribution on the buffer $\mathcal{D}=\{( t, \mathbf{s}_t,\mathbf{a}_t, \mathbf{z}_t)\}$. 
To this end, we build on the NUDGE framework~\cite{nudge}, which jointly optimizes rules via gradient descent and compiles them into deterministic policies at deployment. The discrete distiller consists of three modules: (i) a \textit{spec-grounded vocabulary} $\mathcal{P}$ derived from $\mathcal{C}$, keeping every predicate traceable to protocol specifications; (ii) a \textit{differentiable rule layer} that composes predicates into conjunctions and pools their activations into per-action logits via attention; and (iii) \textit{offline KL distillation} that fits $\hat{\mathbf{z}}_{t,i}$ to the teacher and compiles the rules into a deterministic \texttt{IF-THEN} table.

\textbf{Predicate vocabulary.}
We enumerate a predicate vocabulary $\mathcal{P}$ from the protocol specifications related to $\mathcal{C}$, with $P=|\mathcal{P}|$. Each predicate $p \in \mathcal{P}$ falls into one of two categories:
\begin{itemize}[leftmargin=1.2em,itemsep=0pt,topsep=1pt]
\item \textit{Threshold predicates} (i.e., ``$c_{t,k}$ is either low or high'') for magnitude conditions in KPM-driven policies, e.g., load comparisons as specified in 3GPP.28.552 \cite{3gpp.28.552};
\item \textit{Comparison predicates} (i.e., ``$c_{t,k}$ exceeds $c_{t,k'}$ by offset'') for spec-named measurement events, e.g., the A3/A5 events specified in 3GPP.38.331 \cite{3gpp.38.331}.
\end{itemize}

Boolean-valued predicates have been widely used in classical FOL rules (e.g., $p$ is hard \textit{True} or \textit{False}).
However, it is not effective and informative in our problem. 
Therefore, we introduce a confidence vector, defined as $\mathbf{v}_t = [v_{t, 1}, v_{t, 2}, \dots, v_{t, P}] \in [0,1]^P$, to quantify the predicate valuation.
For example, suppose $\mathcal{P}$ contains  $p_\text{rsrp}$: \texttt{High Target RSRP} and  $p_\text{load}$: \texttt{Low Target Load}; then $v_{t,p_\text{rsrp}}\!=\!0.9$ and $v_{t,p_\text{load}}\!=\!0.2$ indicate 90\% confidence that ``RSRP is very high''  and 20\% confidence that ``the target cell has low load''.
A single affine mapping network parameterized by $\psi=(\mathbf{W}_\psi,\mathbf{b}_\psi)$, with $\mathbf{W}_\psi\in\mathbb{R}^{P\times K}$ and $\mathbf{b}_\psi\in\mathbb{R}^{P}$, maps $\mathbf{c}_t$ to $\mathbf{v}_t$. For each predicate $p\in\mathcal{P}$, let $\mathbf{w}_p\in\mathbb{R}^{K}$ denote the $p$-th row of $\mathbf{W}_\psi$ and $b_p\in\mathbb{R}$ the $p$-th entry of $\mathbf{b}_\psi$. Then, we have
$$v_{t,p} = \sigma\bigl(\mathbf{w}_{p}^{\top} \mathbf{c}_t + b_{p}\bigr),
\label{eq:predicate}$$
where $\mathbf{w}_p$ is masked so each predicate depends only on the concepts it is defined over, and $b_p$ is a learnable threshold.

\textbf{Rule body.} For each action $i$, we associate a finite rule set 
$\mathcal{R}_i$, where each rule $u \in \mathcal{R}_i$ is a conjunction 
of predicates from $\mathcal{P}$ proposing a discrete candidate action 
in $\mathcal{A}_i$ (e.g., $\texttt{IF}\; p \wedge p' \wedge p'' 
\;\texttt{THEN}\; n$, with $p, p', p'' \in \mathcal{P}$ and 
$n \in \mathcal{A}_i$).
This conjunction is realized by the product t-norm:
\begin{equation}
v_{t,u}\;=\;\prod_{p\in u} v_{t,p},
\label{eq:rule_body}
\end{equation}
which transforms the non-differentiable  ``AND'' operator to differentiable ``product'' operator. 
Continuing the above example, the rule $u$: ``\texttt{IF} $p_\text{rsrp}$ \texttt{AND} $p_\text{load}$ \texttt{THEN} handover'' 
yields $v_{t,u} = v_{t,p_\text{rsrp}}\cdot v_{t,p_\text{load}} = 0.18$, 
which means rule $u$ has only a weak impulse of 0.18 to be executed under the current network state. 

\textbf{Action head.}
A rule activation $v_{t,u}$ captures \textit{how strongly} a rule $u$ activates, but not \textit{which candidate action} the rule should trigger. Recall that the teacher's pre-activation vector $\mathbf{z}_{t,i}  \in \mathbb{R}^{d_i}$ has $d_i = |\mathcal{A}_i|$ logits for discrete actions. Let $n \in \{1, \ldots, |\mathcal{A}_i| \}$ index the logits of $\mathbf{z}_{t,i} $. To align rules $u$ with logit $n$, we partition rule set $\mathcal{R}_i$ by the logit, let
\begin{equation}
\mathcal{R}_{i,n} \;=\; \{u \in \mathcal{R}_i : \mathrm{head}(u) = n\}
\end{equation}
collect the rules proposing candidate action logit $n$, so that 
$\mathcal{R}_i = \bigcup_{n \in \mathcal{A}_i} \mathcal{R}_{i,n}$.
\pname{} assigns each rule $u \in \mathcal{R}_i$ a learnable weight 
$w_u \in \mathbb{R}$. Within each subset $\mathcal{R}_{i,n}$, these 
weights are normalized into an \textit{attention distribution} 
\begin{equation}
\alpha_u \;=\; 
\frac{\exp(w_u)}{\sum_{u' \in \mathcal{R}_{i,n}}\exp(w_{u'})},
\qquad u \in \mathcal{R}_{i,n},
\label{eq:rule_attention}
\end{equation}
which is unambiguously defined since each rule belongs to exactly 
one subset. 
Here, $\alpha_u$ is read as the fraction of candidate 
action $n$'s score attributed to rule~$u$. The $n$-th logit of 
$\hat{\mathbf{z}}_{t,i}$ is then the attention-weighted sum of activations 
within that subset:
\begin{equation}
[\hat{z}_{t,i}]_n
=\;\sum_{u \in \mathcal{R}_{i,n}} \alpha_u\cdot v_{t,u}.
\label{eq:action_head}
\end{equation}

\textbf{Distillation training.}
Trained under the KL objective, the 
attention within each $\mathcal{R}_{i,n}$ sharpens onto the rule(s) 
that best explain the teacher's decisions for candidate action~$n$. 
Unlike NUDGE, which updates rule weights via on-policy PPO, \pname
trains the rule layer by offline distillation on~$\mathcal{D}$. With
temperature $\kappa>0$, the teacher and student distributions are jointly optimized by
\begin{equation} \label{eq:dis_loss}
    \mathcal{L}_\text{dis}(\mathbf{W}, \psi) = \frac{1}{N} \sum_{t=1}^{N} D_\text{KL} \left( \text{softmax}(\frac{\mathbf{z}_{t,i} }{ \kappa})
    \parallel \text{softmax}(\frac{\mathbf{\hat{z}}_{t,i} }{\kappa}) \right).
\end{equation}
The distillation can be run asynchronously on an rApp within the non-real-time loop, avoiding resource competition with the tasks within the near-real-time control loop.

\textbf{Compilation to deterministic IF-THEN.}
For each candidate action logit $n\in\mathcal{A}_i$, the dominant rule $u_n^{\star}=\arg\max_{u\in\mathcal{R}_{i,n}}\alpha_u$ is preserved, and rules whose in-subset attention $\alpha_u$ falls below a pruning threshold $W_\text{th}$ are discarded. The result is a plain IF-THEN table, i.e., one deterministic branch per candidate action, that operators can audit line by line and that runs in $O(|\mathcal{R}_i|)$
comparisons on the Near-RT RIC. This compiled program is the student policy~$\pi^{\text{dis}}$.

After per-dimension distillation, \pname concatenates the dimension-wise outputs $\{\hat{z}_{t,i}\}_{i}$ into the full student parameter vector $\hat{\mathbf{z}}_t$, and applies the deterministic head $g(\cdot)$ defined in Eq.~\eqref{eq:sym_mapping} to obtain the candidate student action $\hat{\mathbf{a}}_t = g(\hat{\mathbf{z}}_t)$. Because each dimension is distilled independently from the teacher's factorized marginals, $\hat{\mathbf{a}}_t$ encodes the student's per-dimension intent but does not yet enforce joint operator/spec constraints. It is therefore handed to the symbolic action shield (\S\ref{sec:shield}), which produces the final executed action $\mathbf{\hat{a}}^\star_t$.

\subsection{Symbolic Action Shielding}
\label{sec:shield}

Since the above distillation focuses on behavioral fidelity rather than feasibility, the candidate student action $\hat{\mathbf{a}}_t$ may occasionally violate operator-defined or system-level constraints.  We design a two-stage \emph{symbolic action shielding} mechanism 
to produce the final action $\mathbf{\hat{a}}^\star_t$.
The first stage is \textit{rule-based correction}, which handles constraints expressed in closed form.
The second stage is \textit{safe-decision retrieval}, which handles state-dependent QoS constraints by replaying a previously verified decision. 

\textbf{Rule-based correction.} 
Constraints tractable in closed form, e.g., resource budgets, admissibility sets, per-dimension bounds, are encoded as a finite set of spec-derived correction rules $\tilde{\mathcal{R}}=\{ \tilde{u} \}$. Each rule $\tilde{u}$ couples a closed-form violation function $\nu$ with a closed-form correction operator $\Pi$:
\begin{equation}
\tilde{u}:\texttt{IF} \; \nu(\hat{\mathbf{a}}_t,\mathbf{s}_t)>0 \; \texttt{THEN}\; \hat{\mathbf{a}}_t\leftarrow \Pi(\hat{\mathbf{a}}_t,\mathbf{s}_t),
\end{equation}
where $\nu$ is positive when $(\hat{\mathbf{a}}_t,\mathbf{s}_t)$ violates the constraint encoded by $\tilde{u}$ (e.g., the resource budget is exceeded), and $\Pi$ projects the action back into the feasible region, for instance, via simplex renormalization, clamping to per-entry bounds, or rounding to the nearest admissible value. Starting from the candidate $\hat{\mathbf{a}}_t$, rules are applied in spec-prioritized order; the result is denoted as $\hat{\mathbf{a}}'_t$.

\textbf{Safe-decision retrieval.}
Rule-based correction handles only constraints expressible in closed form and acts pointwise on the candidate $\hat{\mathbf{a}}_t$. The corrected action $\hat{\mathbf{a}}'_t$ can still violate state-dependent QoS constraints introduced in \S\ref{sec:system-model}, e.g., per-slice throughput or delay targets, whose feasibility depends on how the radio environment subsequently evolves under $\hat{\mathbf{a}}'_t$ and therefore admits no static projection. \pname closes this gap by replaying past decisions that have already been observed to satisfy QoS in operationally similar contexts, rather than synthesizing a new feasible action online. 

\pname maintains an online \textit{safe-decision bank} as
\begin{equation}
    \mathcal{D}_\text{safe} = \bigl\{(\mathbf{c}_j, \mathbf{a}^\star_j)\bigr\},
\end{equation}
collecting past executed actions whose subsequently observed $T'$-step QoS-violation cost $\Lambda(\mathbf{s}_t, \mathbf{\hat{a}}^\star_t)$ stayed within operator operator-tolerated bounds, with $T' \le T$. Entries are collected only after post-hoc verification, so every element of $\mathcal{D}_\text{safe}$ is known-safe in the context in which it was executed. A lightweight risk estimator $q:[0,1]^K\!\times\!\mathcal{A}\to[0,1]$, instantiated as a two-layer MLP, scores the constraint risk of $\hat{\mathbf{a}}'_t$ and acts \emph{only} as a QoS-violation gate. We train $q$ offline on logged triples $(\mathbf{c}_t,\mathbf{a}_t,y_t)$, where $y_t\!\in\!\{0,1\}$ flags whether $\Lambda$ stayed within tolerance over the subsequent $T'$ steps. Once $q$ is trained, we record the empirical distribution of its scores $\{q(\mathbf{c}_j,\mathbf{a}^\star_j)\}_{j\in\mathcal{D}_\text{safe}}$ on the training buffer and fix the trigger threshold $\delta$ based on the distribution of the score. Retrieval then executes when $q(\mathbf{c}_t,\hat{\mathbf{a}}'_t) > \delta$.

Upon this trigger, the shield replaces $\hat{\mathbf{a}}'_t$ with the
known-safe entry whose concept vector is closest to the current one,
\begin{equation}
k^\star = \arg\min_{k:\,(\mathbf{c}_k,\cdot)\in\mathcal{D}_\text{safe}}
\|\mathbf{c}_t - \mathbf{c}_k\|_2,
\qquad \mathbf{\hat{a}}^\star_t \leftarrow \hat{\mathbf{a}}_{k^\star}. 
\label{eq:retrieve}
\end{equation}
Proximity is measured in concept space rather than in raw state space, so the metric is low-dimensional and the $O(|\mathcal{D}_\text{safe}|\!\cdot\!K)$ query is cheap. Otherwise (when $q(\mathbf{c}_t,\hat{\mathbf{a}}'_t)\le\delta$), the rule-corrected action passes through directly: $\mathbf{\hat{a}}^\star_t \leftarrow \hat{\mathbf{a}}'_t$.




\section{Experiments}
\label{Experiment_5}

We evaluate \pname on two representative O-RAN control tasks:
(i) resource slicing representing continuous action space (\S\ref{sec:uc_slicing}), and (ii) UE handover representing discrete action space (\S\ref{sec:uc_ho}). Both tasks are evaluated on a live indoor multi-cell 5G~NR testbed shown in Fig.~\ref{fig:oran_eb3}. 
The evaluation seeks to answer four questions:
\begin{itemize}[leftmargin=1.2em,itemsep=-1pt,topsep=1pt]
    \item \textbf{Q1~Interpretability.}
    Are the concept and the student policy spec-auditable, and how simple are they relative to SOTA XAI baselines under a complexity metric $\Omega$?
    \item \textbf{Q2~Control performance.}
    Does \pname preserve the teacher's reward and QoS guarantees on live
    5G traffic?
    \item \textbf{Q3~System overhead.}
    Does \pname satisfy the Near-RT RIC latency and memory requirements?
    \item \textbf{Q4~Component contribution.}
    How does each \pname module (\textit{conceptizer}, \textit{symbolic distiller}, \textit{action shield}) contribute to the end-to-end result?
\end{itemize}

\subsection{5G NR O-RAN Testbed}
\label{sec:testbed}

\begin{figure}[t]
    \centering
    \includegraphics[width=0.98\linewidth]{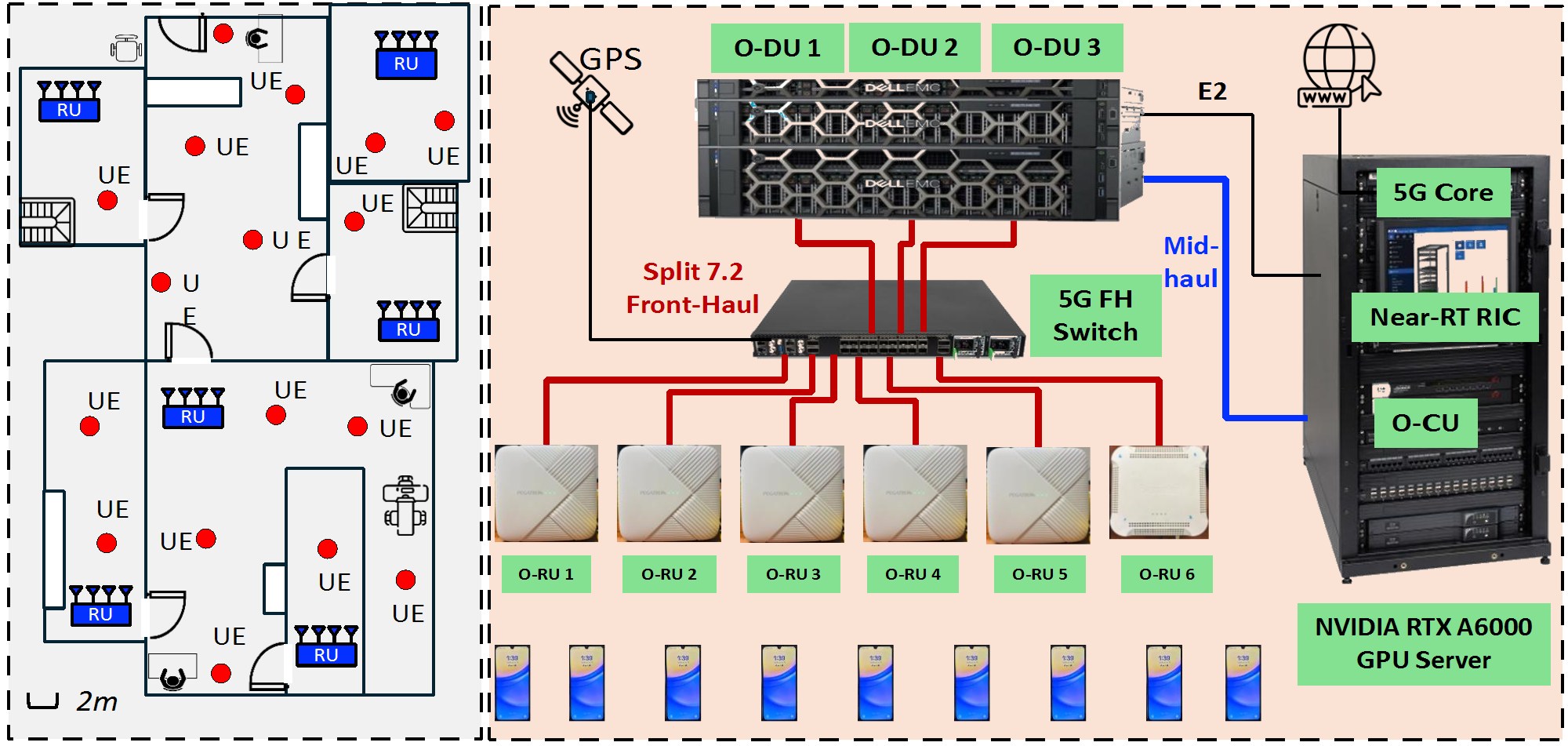}\vspace{-0.1in}
    \caption{A multi-cell indoor 5G NR O-RAN testbed.}\vspace{-0.1in}
    \label{fig:oran_eb3}
\end{figure}

Fig.~\ref{fig:oran_eb3} shows our 5G NR O-RAN testbed deployed within a building. The protocol stack is built on srsRAN~25.10~\cite{srsran_project} with Split~7.2 fronthaul and Split~Option~2 midhaul. Commercial RUs from Benetel and Pegatron are integrated into the deployment, operating on the 5G n78 band with $100$~MHz bandwidth and $4\times4$ MIMO. The core network is implemented on Open5GS. To emulate realistic network dynamics, the testbed incorporates $14$ smartphones of mixed vendors (OnePlus~Nord~AC2003, Motorola~G54, Samsung~Galaxy~A15). Traffic is generated via \texttt{iperf} to produce continuous and heterogeneous flows across three network slices with time-varying loads. The ORAN-SC Near-RT RIC~\cite{oran-sc-ric} is containerized and deployed on a server with NVIDIA RTX PRO 6000 Blackwell GPU. \pname's fast-loop (conceptizer inference and shield) runs inside an xApp co-located with the RIC, while the slow-loop (conceptizer and symbolic distiller training) runs asynchronously on an rApp host.

GitHub link to the source code will be provided if the paper is accepted for publication.


\subsection{Performance Metrics and Baselines}
\label{sec:protocol}

\textbf{Baselines.}
We compare \pname against its DRL teacher and two SOTA XAI baselines, Metis~\cite{metis} and SYMBXRL~\cite{symbxrl}. To isolate the effect of the distillation method from that of the teacher, both baselines are trained the same teacher traces as \pname from the buffer $\mathcal{D}$; once distilled, every student policy is frozen and deployed on the testbed for online evaluation without further teacher involvement.
\begin{itemize}[leftmargin=1.2em,itemsep=1pt,topsep=2pt]
    \item \textbf{Teacher.} We use Proximal Policy Optimization (PPO) for continuous resource slicing and Double DQN for discrete UE handover.
    They serve as the performance upper bound.
    \item \textbf{Metis}~\cite{metis}. We use its decision-tree mode, which fits a regression tree or a classification tree (depending on the task type) directly to the teacher's trace $(\mathbf{s}_t,\mathbf{a}_t)$ over the raw KPMs.
    \item \textbf{SYMBXRL}~\cite{symbxrl}. This approach discretizes the continuous states into first-order-logic (FOL) predicates and constructs a knowledge graph over the predicates and actions.
\end{itemize}

\textbf{Interpretability metric $\Omega$.}
We quantify the complexity of a policy by $\Omega$, defined as the number of atoms in its symbolic representation. An atom is the smallest indivisible syntactic unit of the representation: nodes of the expression tree (operators, variables, constants) for mathematical expressions; logical connectives, comparison operators, predicates, variables, and constants for FOL rules; nodes for decision trees; subjects, relations and objects in the knowledge graphs. We report $\Omega$ jointly with the mean step reward as a Pareto curve.

\textbf{Training and reporting.}
Each baseline is evaluated across five independent runs of $10^4$ decision steps, with each step triggered by a KPM report from the E2 node every $500$~ms.

\subsection{Use Case 1: Resource Slicing}
\label{sec:uc_slicing}

\subsubsection{Task Formulation}
\label{sec:uc1_form}

We consider a case where a cell serves a set of time-varying UEs. 
Here, $\mathcal{G}_t$ is the set of UEs, with $|\mathcal{G}_t|\in[9,14]$.
The UEs are partitioned into three disjoint slice-level subsets so that $\mathcal{G}_t=\mathcal{G}_\text{eMBB}\cup\mathcal{G}_\text{URLLC}\cup\mathcal{G}_\text{mMTC}$. 
Denote $N_\text{PRB}$ as the total number of physical resource blocks (PRBs).
The state $\mathbf{s}_t$ is the per-UE KPM row $\{\mathbf{s}_{t,g}\}_{g\in\mathcal{G}_t}$ of Table~\ref{tab:kpm}, and the action 
$\mathbf{a}_t=(a_{t,\text{eMBB}},a_{t,\text{URLLC}},a_{t,\text{mMTC}})$ is a continuous per-slice PRB ratio on the simplex. The PPO teacher maximizes
\begin{equation}
r_t^\text{rs}=\sum_{g\in\mathcal{G}_t}\!
\Big[ \underbrace{\tfrac{1}{1+\sum_{g\in\mathcal{G}_t} n_{t,g}}}_{\text{Resource efficiency}} - \underbrace{\left( \beta_1 r_{t,g}^{[p]} + \beta_2 r_{t,g}^{[d]} \right)}_{\text{QoS penalty}} \Big].
\label{eq:reward_rs}
\end{equation}
where $n_{t,g}$ is the PRBs allocated to UE $g$, $r_{t,g}^{[p]} = \max ( (\tilde{\mu}_{t,g} - {\mu}_{t,g})/\tilde{\mu}_{t,g},0 )$, and $r_{t,g}^{[d]} = \max ( ({\omega}_{t,g} - \tilde{\omega}_{t,g})/\tilde{\omega}_{t,g},0 )$ are the QoS penalties, where $\mu_{t,g},\omega_{t,g}$ are the achieved throughput and delay, $(\tilde{\mu}_{t,g},\tilde{\omega}_{t,g})$ are the corresponding QoS targets, weighted by $\beta_1, \beta_2 > 0$. 

In addition, since the throughput and delay demands vary across slices, we report the per-step violation rates
$V_\text{thp}(t)=|\mathcal{G}_t|^{-1}\!\sum_{g\in\mathcal{G}_t}\mathbf{1}(\mu_{t,g}\!<\!\tilde{\mu}_{t,g})$ and
$V_\text{dly}(t)=|\mathcal{G}_t|^{-1}\!\sum_{g\in\mathcal{G}_t}\mathbf{1}(\omega_{t,g}\!>\!\tilde{\omega}_{t,g})$. The concept scheme $\mathcal{C}_\text{RS}$ is in the Table~\ref{tab:semantic_mapping}, i.e., eMBB demand, URLLC stress, slice load, channel quality.

\subsubsection{Interpretability}
\label{sec:uc1_interp}

\begin{figure}[t]
    \centering
    \begin{minipage}{0.52\linewidth}
        \centering
        \includegraphics[width=\linewidth]{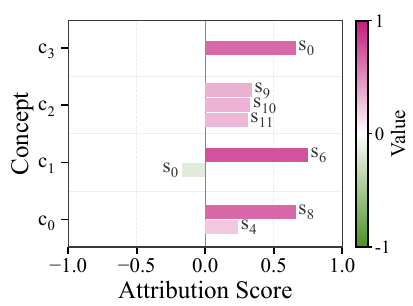}\vspace{-0.1in}
        \caption{IG attribution from KPMs to slicing concepts.}
        \label{fig:rs_attr}
    \end{minipage}
    \hfill
    \begin{minipage}{0.43\linewidth}
        \centering
        \includegraphics[width=\linewidth]{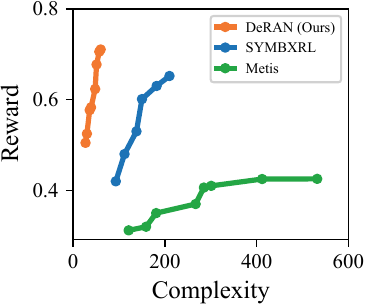}\vspace{-0.05in}
        \caption{Reward vs.\ symbolic complexity.}
        \label{fig:rs_pareto}
    \end{minipage}\vspace{-0.15in}
\end{figure}


\textbf{Concept auditing.}
Fig.~\ref{fig:rs_attr} shows the normalized IG values of each KPM for each concept over the aggregated UEs. For every concept $c_k\in\mathcal{C}_\text{RS}$, the attribution concentrates inside the support $\mathcal{M}_k$ in by Table~\ref{tab:semantic_mapping}: $c_0$ (eMBB demand) loads on $s_4,s_8$; $c_1$ (URLLC stress) on $s_0,s_6$;
$c_2$ (slice load) on $s_9,s_{10},s_{11}$; and $c_3$ (channel quality) on $s_0$. 
This confirms that the per-concept masking in \S\ref{sec:conceptizer} eliminates off-support leakage by construction. The sign pattern also agrees with the semantic names: $s_6$ (BLER) drives $c_1$ (URLLC stress) positively while $s_0$ (CQI) drives it negatively, consistent with ``high stress = poor channel + high error rate.''

\begin{figure}[t]
\centering
\begin{tcolorbox}[
  enhanced,
  colback = gray!8,
  colframe = gray!55!black,
  boxrule = 0.4pt,
  arc = 1.20mm,
  left = 0pt, 
  right = 0pt, 
  top = -5pt, 
  bottom = 1pt,
  boxsep = 1pt,
  width = 0.95\linewidth
]
\footnotesize
\begin{align*}
z_{\text{eMBB}}  &= \log(1+c_{0}) + 1.32\,c_{3} + 0.62\,c_{0}c_{3} - 1.86\,c_{2} - 0.45\,c_{1}, \\
z_{\text{URLLC}} &= 1.74\,c_{1} + 1.26\,c_{1}^{2} + 0.62\,c_{1}c_{2} - 0.74\,c_{3}, \\
z_{\text{mMTC}}  &= \frac{1}{1.42 + c_{0} + c_{1}} - 0.36\log(1+c_{2}) + 0.23\,c_{3}.
\end{align*}
\end{tcolorbox}\vspace{-0.1in}
\caption{An instance of the distilled formulas for slicing.}
\label{fig:formula}
\end{figure}

\textbf{Complexity.} Fig.~\ref{fig:rs_pareto} reports the reward-complexity Pareto comparison of the three XAI methods under a sweep of hyperparameters (number of leaves for Metis, KPM change threshold for SYMBXRL, expression-tree depth for \pname. At its highest-reward operating point, \pname achieves mean step reward $0.71$ at $\Omega{=}56$, while SYMBXRL requires $\Omega{=}206$ to reach $0.65$ and Metis saturates at $0.43$ even at $\Omega{=}532$. 
Compared to SYMBXRL, \pname achieves $1.09\times$ higher reward with $3.7\times$ fewer nodes; relative to Metis, the gap widens to $1.65\times$ reward at $9.5\times$ lower complexity. The gap is driven by the concept bottleneck: \pname distills over four operator-named variables, whereas Metis and SYMBXRL must recover the same structure from $|\mathcal{M}|=12$ KPMs.

\textbf{Symbolic policy auditing.} 
The distilled \pname student for each slice is a single closed-form expression $\hat{z}_{t,i}=\pi^\text{con}_i(\mathbf{c}_t)$ over four concepts, which we provide in Fig.~\ref{fig:formula}.
Each expression is operationally readable: the eMBB logit rises with eMBB demand ($\log (1+c_0)$) and channel quality ($c_3$), and is penalized by competing slice load ($c_2$) and URLLC stress ($c_1$); the URLLC logit grows super-linearly in $c_1$ (a near-quadratic reflex to rising URLLC stress) and is inversely coupled to channel quality; the mMTC logit is a residual claim on PRBs that relaxes when either eMBB or URLLC demand grows.

\subsubsection{Control Performance}
\label{sec:uc1_ctrl}

\begin{figure}[t]
    \centering
    \begin{subfigure}[t]{0.325\linewidth}
        \centering
        \includegraphics[width=\linewidth]{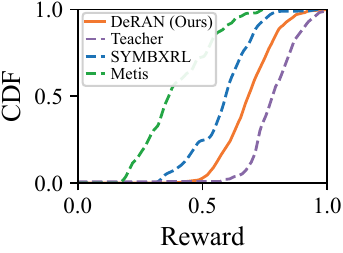}
        \caption{CDF of reward.}
        \label{fig:rs_reward}
    \end{subfigure}
    \hfill
    \begin{subfigure}[t]{0.325\linewidth}
        \centering
        \includegraphics[width=\linewidth]{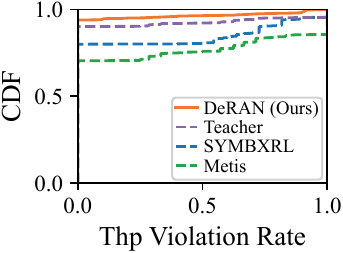}
        \caption{CDF of throughput violation rate $V_\text{thp}$.}
        \label{fig:rs_thp}
    \end{subfigure}
    \hfill
    \begin{subfigure}[t]{0.325\linewidth}
        \centering
        \includegraphics[width=\linewidth]{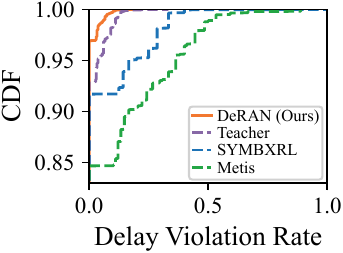}
        \caption{CDF of delay violation rate $V_\text{dly}$.}
        \label{fig:rs_dly}
    \end{subfigure}\vspace{-0.1in}
    \caption{Control performance comparison on the resource-slicing task.}\vspace{-0.15in}
    \label{fig:rs_ctrl}
\end{figure}

Fig.~\ref{fig:rs_ctrl} reports the CDF of reward together with $V_\text{thp}$ and $V_\text{dly}$ for \pname, Metis, SYMBXRL, and the teacher over five independent runs.

\textbf{Reward}. In Fig.~\ref{fig:rs_reward}, \pname follows the teacher closely across the entire reward distribution: the median reward of \pname is $0.65$ against the teacher's $0.75$ (a $87\%$ recovery), and the lower tail at CDF$=0.1$ is within $0.02$ of the teacher. SYMBXRL lags the teacher by $0.20$ at the median ($73\%$ recovery), and Metis collapses to a $0.38$ median ($51\%$ recovery) because the hard splits in the tree generation cannot express the smooth multi-slice trade-off that the teacher learns.

\textbf{QoS}. \pname is the only student that outperforms the teacher. Fig.~\ref{fig:rs_thp} shows that $95\%$ of \pname's decision steps  maintain $V_\text{thp}\approx 0$, against $90\%$ for the teacher, $80\%$ for SYMBXRL, and $68\%$ for Metis; the crossover between \pname and the teacher at the low tail is produced by the safe-decision retrieval stage of the action shielding mechanism (\S\ref{sec:shield}), which replaces the candidate action whenever the risk estimator flags throughput risk. The delay-violation distribution in Fig.~\ref{fig:rs_dly} follows the same ordering: \pname ($97\%$ near-zero) $\approx$ teacher ($94\%$) $>$ SYMBXRL ($92\%$) $>$ Metis ($83\%$). Aggregated across all steps, \pname reduces the mean $V_\text{thp}$ by $3.9\times$ versus SYMBXRL and $5.8\times$ versus Metis, and the mean $V_\text{dly}$ by $2.3\times$ and $5.1\times$ respectively.

\subsubsection{System Overhead}
\label{sec:uc1_sys}

\begin{table}[t]
\centering
\caption{System overhead on the resource-slicing task.}\vspace{-0.1in}
\label{tab:sys_rs}
\resizebox{\columnwidth}{!}{%
\begin{tabular}{lccccc}
\toprule
\textbf{Method} & \!\!\!\textbf{P50 latency}~($\mu$s)$\downarrow$\! & \!\textbf{P99 latency}~($\mu$s)$\downarrow$\!\!\!
    & \textbf{max}~($\mu$s)~$\downarrow$ & \!\!\textbf{RSS}~(MB)~$\downarrow$
    & $\boldsymbol{\Omega}$~$\downarrow$  \\
\midrule
Teacher      & 2{,}100 & 3{,}800 & 7{,}200 & 410 & ---      \\
Metis               & 45      & 110     & 340     & 12  & 532     \\
SYMBXRL             & 380     & 950     & 2{,}100 & 85  & 206      \\
\textbf{\pname (Ours)\!\!\!\!}                    & 95      & 265     & 520     & 18  & 56       \\
\bottomrule
\end{tabular}%
}
\end{table}
Table~\ref{tab:sys_rs} compares the 50th-percentile (P50) lantecy, 99th-percentile (P99) latency, and the memory footprint measured by peak resident-set size (RSS) of the teacher, Metis, SYMBXRL, and \pname. The  teacher operates on raw KPMs through a DNN forward pass, while \pname inference on the concept vector $\mathbf{c}_t$ followed by a per-dimension closed-form expression on $\mathbf{c}_t$. As a result, \pname reduces the 99th-percentile latency from $3{,}800\,\mu\text{s}$ (teacher) to $265\,\mu\text{s}$ (a $14.3\times$ speedup) and the RSS from $410$\,MB to $18$\,MB ($22.8\times$). Compared to SYMBXRL, which must traverse a knowledge graph at every step, \pname is $3.6\times$ faster in terms of the 99th-percentile latency. Although Metis yields the lowest P99 latency (110 $\mu$s), it comes at the expense of a $9.5\times$ larger $\Omega{=}532$ overhead and a severe performance drop to a  $0.38$ median reward. Overall, \pname achieves a Pareto improvement over all baselines regarding the reward-latency trade-off.

\subsubsection{Ablation Studies}
\label{sec:uc1_abl}


\begin{table}[t]
\centering
\caption{Ablation study of \pname on the resource-slicing task. Reward is normalized w.r.t. the teacher; $V_\text{thp}$ and $V_\text{dly}$ are calculated for per-UE step violation rates.}\vspace{-0.1in}
\label{tab:ablation_rs}
\resizebox{\columnwidth}{!}{%
\begin{tabular}{llcccc}
\toprule
\textbf{Variant} & \textbf{Option} & \textbf{Reward}~$\uparrow$
    & $\boldsymbol{V_\text{thp}}$\%~$\downarrow$ & $\boldsymbol{V_\text{dly}}$\%~$\downarrow$
    & $\boldsymbol{\Omega}$~$\downarrow$  \\
\midrule
Teacher & -- & 1.000    & 7.4  & 1.5  & ---        \\
\midrule 
\pname (full) & -- & 0.782  &  3.6  & 0.6  & 56          \\
\midrule

\multirow{2}{*}{Conceptizer} 
& w/o conceptizer (DSR)        & 0.583  & 15.2 & 12.4 & 78          \\
& w/o per-concept masking      & 0.755  & 8.1  & 7.3  & 62          \\
\midrule

\multirow{2}{*}{Distiller} 
& DSR $\rightarrow$ linear on $\mathbf{c}_t$ & 0.743  & 11.2 & 9.8  & 34        \\
& DSR $\rightarrow$ MLP on $\mathbf{c}_t$    & 0.868  & 2.9  & 1.8  & 3{,}217  \\
\midrule

\multirow{3}{*}{Shield} 
& w/o rule-based correction    & 0.763  & 12.0 & 5.9  & 56          \\
& w/o safe-decision retrieval  & 0.752  & 6.8  & 10.7 & 56        \\
& w/o both shields             & 0.622  & 17.6 & 15.8 & 56        \\
\bottomrule
\end{tabular}%
}
\end{table}

Table~\ref{tab:ablation_rs} reports the contribution of each \pname
module on the slicing task. Three findings stand out. \textit{(i)~The conceptizer is the dominant driver of reward performance}: replacing it with DSR directly over the raw KPMs drops the reward from $0.782$ to $0.583$ and increases $V_\text{thp}$ and $V_\text{dly}$ by $4.2\times$ and $20.7\times$ respectively, because DSR on a $\sim$12-dimensional KPM vector exhausts its search budget. Removing only the per-concept masking (but keeping the concept scheme) costs $0.03$ reward but doubles $V_\text{thp}$. \textit{(ii)~DSR outperforms linear distillation and achieves comparable performance to a MLP with lower complexity}: replacing DSR by a linear head on $\mathbf{c}_t$ loses $0.04$ reward, while replacing it by a dense MLP recovers $0.086$ reward over full \pname but increases $\Omega$ by $57\times$, eliminating auditability. \textit{(iii)~The two shield stages are complementary}: removing rule-based correction primarily damages $V_\text{thp}$ (driven by simplex violations), while removing safe-decision retrieval primarily damages $V_\text{dly}$ (driven by state-dependent QoS risk); removing both is strictly worse than either ablation and brings the student close to an un-shielded symbolic controller.

\subsection{Use Case 2: UE Handover}
\label{sec:uc_ho}

\subsubsection{Task Formulation}
\label{sec:uc2_form}

A target UE $g$ in the overlap of two cells $\mathcal{G}=\{\text{serv},\text{tgt}\}$ decides each step to stay on the serving cell ($a_t{=}0$) or switch to the target cell ($a_t{=}1$), with $\mathbf{z}_t\in\mathbb{R}^{2}$ holding the corresponding Q-values. The state KPMs $\mathbf{s}_t$ are shown in Table~\ref{tab:kpm}. The  teacher maximizes
\begin{equation}
r_t^\text{ho}= - \underbrace{\left( \beta_3 r_{t,g}^{[p]} + \beta_4 r_{t,g}^{[d]} \right)}_{\text{QoS penalty}},
\label{eq:reward_ho}
\end{equation}
where $\beta_3,\beta_4 > 0$ are weights, and the two penalty terms for throughput and delay deviations are defined identically to \eqref{eq:reward_rs}. The concept scheme $\mathcal{C}_\text{HO}$ is in the Table~\ref{tab:semantic_mapping}, i.e., serving/target signal quality and load ($c_4{-}c_7$) and QoS-degradation status ($c_8$), with $|\mathcal{G}_k|{=}1$ for the signal-quality and load concepts.

\subsubsection{Interpretability}
\label{sec:uc2_interp}

\begin{figure}[t]
    \centering
    \begin{minipage}{0.5\linewidth}
        \centering
        \includegraphics[width=\linewidth]{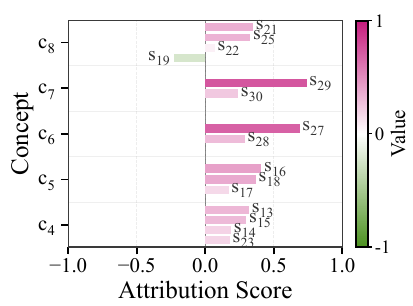}
        \caption{IG attribution from KPMs to handover concepts.}
        \label{fig:ho_attr}
    \end{minipage}
    \hfill
    \begin{minipage}{0.43\linewidth}
        \centering
        \includegraphics[width=\linewidth]{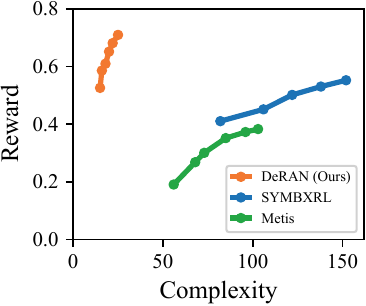}
        \caption{Reward vs.\ symbolic complexity.}
        \label{fig:ho_pareto}
    \end{minipage}
\end{figure}

\begin{figure}[t]
\centering
\vspace{-0.15in}
\begin{tcolorbox}[
  enhanced,
  colback = gray!8,            
  colframe = gray!55!black,    
  boxrule = 0.4pt,
  arc = 1.2mm,
  left = 4pt, right = 4pt, top = 4pt, bottom = 4pt,
  boxsep = 2pt,
  width = 0.7\linewidth          
]
\scriptsize  
\begin{algorithmic}
\State \textbf{if} $c_5 - c_4 > 0.12$ \textbf{and} $c_7 < 0.47$
\textbf{then} 
\State \quad \textbf{switch}
\State \textbf{else if} $c_4 < 0.32$ \textbf{and} $c_5 > 0.58$ 
\textbf{then} 
\State \quad \textbf{switch}
\State \textbf{else if} $c_8 > 0.50$ \textbf{and} $c_5 > 0.63$ \textbf{and} $c_7 < 0.42$ 
\textbf{then} 
\State \quad \textbf{switch}
\State \textbf{else}
\State \quad \textbf{stay}
\State \textbf{end if}
\end{algorithmic}
\end{tcolorbox}\vspace{-0.15in}
\caption{An instance of the distilled rules for handover task.}\vspace{-0.15in}
\label{fig:decision_code}
\end{figure}

\textbf{Concept auditing.}
Fig.~\ref{fig:ho_attr} reports the IG attribution of the five handover concepts across the handover KPM set in Table~\ref{tab:kpm}. The attribution pattern matches the semantic names: $c_4$ (serving signal quality) loads on $\{s_{13-15},s_{23}\}$ (serving RSRP/RSRQ/SINR/CQI), $c_5$ (target signal quality) on $\{s_{16},s_{17},s_{18}\}$, $c_6,c_7$ split onto serving- and target-side PRB utilization ($s_{27},s_{28}$ vs $s_{29},s_{30}$), and $c_8$ (QoS degradation) loads positively on BLER/delay-burst KPMs while loading \emph{negatively} on $s_{19}$ (achieved throughput), i.e., degradation grows when throughput drops. Crucially, the serving- and target-cell supports of $c_4$ and $c_5$ are \emph{disjoint} in the IG map: no KPM from the target cell contaminates the serving-cell concept or vice versa, which is a direct consequence of the per-concept masking of \S\ref{sec:conceptizer}. 

\textbf{Complexity.} Fig.~\ref{fig:ho_pareto} plots the reward-complexity Pareto for the three XAI methods. \pname's highest operating point is $(\Omega{=}26,\text{reward}{=}0.71)$, against $(\Omega{=}152,\text{reward}{=}0.55 )$ for SYMBXRL and $(\Omega{=}103,\text{reward}{=}0.38)$ for Metis. Relative to SYMBXRL, \pname achieves $1.29\times$ higher reward with $6.0\times$ fewer nodes; relative to Metis, $1.87\times$ reward at $4.0\times$ lower complexity. The gap is again attributable to the concept bottleneck: the distiller reasons over five named concepts rather than the 19 raw KPMs available to the baselines, so the higher reward target is reachable with far simpler predicates.

\textbf{Symbolic policy auditing.}
The compiled \pname student (Fig.~\ref{fig:decision_code}) is a four-branch IF-THEN table in which every threshold is a concept comparison (rather than a raw KPM). The first branch encodes the canonical A3 semantics at the \emph{concept} level (``target signal exceeds serving by a margin \emph{and} target is not overloaded''); the second branch triggers a rescue handover when the serving cell has degraded in absolute terms; the third branch binds mobility to QoS pressure, firing only when $c_8$ is elevated and the target is both strong and lightly loaded. The table runs in four comparisons and is line-by-line auditable against the 3GPP~TS~38.331 A3/A5 vocabulary.

\subsubsection{Control Performance}
\label{sec:uc2_ctrl}

\begin{figure}[t]
    \centering
    \begin{subfigure}[b]{0.49\columnwidth}
        \centering
        \includegraphics[width=\textwidth]{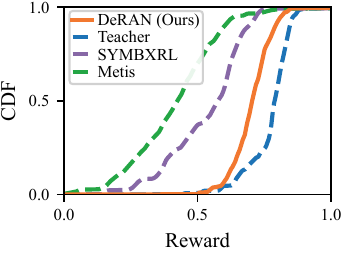}
        \caption{CDF of reward.}
        \label{fig:ho_reward}
    \end{subfigure}
    \hfill
    \begin{subfigure}[b]{0.49\columnwidth}
        \centering
        \includegraphics[width=\textwidth]{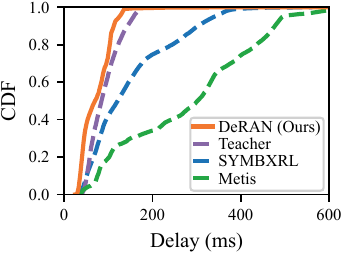}
        \caption{CDF of delay.}
        \label{fig:ho_dly}
    \end{subfigure}\vspace{-0.1in}
    \caption{Control performance on the UE-handover task.}\vspace{-0.1in}
    \label{fig:ho_ctrl}
\end{figure}

Fig.~\ref{fig:ho_ctrl} reports the per-step reward and delay CDFs for
\pname, the teacher, Metis, SYMBXRL.

\textbf{Reward}. \pname demonstrates competitive performance to the teacher : the median reward is $0.70$ against the teacher's $0.76$, a $92\%$ recovery. SYMBXRL and Metis underperforms at medians of $0.55$ and $0.45$.

\textbf{QoS}. \pname outperforms the three baselines in terms of the delay: the median per-step delay of \pname is $\sim$$50$\,ms, against $\sim$$80$\,ms for the teacher, $\sim$$150$\,ms for SYMBXRL, and $\sim$$350$\,ms for Metis. At P99, \pname reaches $140$\,ms, improving the teacher by $1.2\times$, SYMBXRL by $2.9\times$, and Metis by $4.1\times$. The performance improvement over the teacher policy comes from the action shielding stage. This mechanism replays a previously verified handover whenever the risk estimator is triggered due to potential ping-pong or congestion-driven delay spike, effectively mitigating the tail of the delay.

\subsubsection{System Overhead}
\label{sec:uc2_sys}

\begin{table}[t]
\centering
\caption{System overhead on the UE-handover task.}\vspace{-0.1in}
\label{tab:sys_ho}
\resizebox{\columnwidth}{!}{%
\begin{tabular}{lccccc}
\toprule
\textbf{Method} & \!\!\!\textbf{P50 latency}~($\mu$s)$\downarrow$\! & \!\textbf{P99 latency}~($\mu$s)$\downarrow$\!\!\!
    & \textbf{max}~($\mu$s)~$\downarrow$ & \!\!\textbf{RSS}~(MB)~$\downarrow$
    & $\boldsymbol{\Omega}$~$\downarrow$  \\
\midrule
Teacher      & 1{,}800 & 3{,}100 & 5{,}400 & 180 & ---  \\
Metis               & 30      & 70      & 180     & 9   & 103   \\
SYMBXRL             & 290     & 820     & 1{,}800 & 75  & 152  \\
\textbf{\pname (Ours)}        & 52      & 194     & 420     & 15  & 25   \\
\bottomrule
\end{tabular}%
}\vspace{-0.15in}
\end{table}

As shown in Table~\ref{tab:sys_ho}, \pname achieves a P99 latency of 194$\mu$s and a peak RSS of 15 MB. Compared to the teacher policy's 3,100 $\mu$s and 180 MB, this represents a 16$\times$ reduction in tail latency and a 12$\times$ decrease in memory footprint. Furthermore, avoiding expensive knowledge-graph traversals makes \pname 4.2$\times$ faster than SYMBXRL at the 99th percentile. Although Metis evaluates faster (70 $\mu$s), it comes at a 4.1$\times$ higher structural complexity ($\Omega=103$ vs. 25), as analyzed in \S\ref{sec:uc2_interp}. Ultimately, while all XAI students satisfy the Near-RT processing budget, \pname is the only method that uniquely strikes the optimal balance: it combines sub-millisecond execution and the lowest structural complexity ($\Omega=25$) while maintaining 87\% of the teacher's reward.

\subsubsection{Ablation Studies}
\label{sec:uc2_abl}


\begin{table}[t]
\centering
\caption{Ablation study of \pname on the UE-handover task. }\vspace{-0.1in}
\label{tab:ablation_ho}
\resizebox{\columnwidth}{!}{%
\begin{tabular}{llccc}
\toprule
\textbf{Variant} & \textbf{Option} & \textbf{Reward}~$\uparrow$
    & \textbf{P99 Delay} (ms)~$\downarrow$
    & $\boldsymbol{\Omega}$~$\downarrow$  \\
\midrule
Teacher & -- & 1.000    & 172    & ---   \\
\midrule
\pname (full) & -- & 0.871  & 135   & 26    \\
\midrule

\multirow{2}{*}{Conceptizer} 
& w/o conceptizer (raw-KPM)    & 0.543  & 388  & 44    \\
& w/o per-concept masking      & 0.782  & 342   & 38     \\
\midrule

\multirow{2}{*}{Distiller} 
& NUDGE $\rightarrow$ linear on $\mathbf{c}_t$ & 0.794 & 172 & 30\\
& NUDGE $\rightarrow$ MLP on $\mathbf{c}_t$    & 0.868  & 148  & 2{,}105  \\
\midrule

Shield 
& w/o safe-decision retrieval  & 0.805 & 265    & 26   \\
\bottomrule
\end{tabular}%
}
\end{table}

We evaluate the same components as in the slicing scenario, but exclude the simplex correction because it is inapplicable to binary handover actions. 
Table~\ref{tab:ablation_ho} presents our experimental results. 
We observe two findings consistent with the slicing case and one unique to the handover: \textit{(i)~The conceptizer is the primary contributor to the reward performance}. Replacing it with raw-KPM predicates degrades the reward from 0.871 to 0.543 and increases the P99 delay from 135 ms to 388 ms. This degradation occurs because NUDGE's rule search suffers from state-space explosion in the raw-KPM space and fails to effectively distinguish between serving and target cells. Furthermore, per-concept masking alone yields a 0.09 reward improvement and a 2.5$\times$ reduction in P99 delay. \textit{(ii) NUDGE significantly outperforms linear distillation}. It achieves a reward within 0.003 of an MLP model while maintaining an 81$\times$ lower $\Omega$. This confirms that the IF-THEN rules are sufficiently expressive for handovers when grounded in high-level concepts. \textit{(iii) Safe-decision retrieval accounts for nearly the entire tail-delay reduction.} Omitting this module maintains a relatively high reward (0.805) but degrades the P99 delay to 265 ms, as the distilled IF-THEN rules occasionally trigger ping-pong handovers under extreme load conditions.


\section{Related Work}
\label{Related_Work_6}

\textbf{DRL for Network Automation.}
DRL has become a key driver of network control and management in O-RAN. For example, ORANSlice \cite{Oranslice} demonstrates DRL-based xApps for closed-loop resource slicing, while xSlice \cite{yan2025near} and EExApp \cite{eexapp} incorporate GNNs and multi-actor PPO for topology-aware slicing and energy-efficient RU scheduling. In mobility and system optimization, prior work explores hierarchical multi-agent DRL for handover \cite{hmarl} and frameworks such as Mutant \cite{mutant} and DTOTO \cite{drl_mec} for congestion control and MEC offloading. Despite these advances, existing approaches rely on black-box neural networks with limited interpretability and high computational overhead, hindering efficient deployment on resource-constrained platforms.

\newcommand{\cmark}{\ding{51}} 
\newcommand{\xmark}{\ding{55}} 
\newcommand{\pmark}{$\circ$}   

\begin{table}[!t]
\centering
\begin{threeparttable}
\caption{Comparison of existing XAI work for O-RAN.} 
\label{tab:xai_comparison}
\footnotesize
\setlength{\tabcolsep}{3.5pt}
\renewcommand{\arraystretch}{1.05}
\begin{tabular}{@{}l c c c c c@{}}
\toprule
\textbf{Work} & \textbf{Paradigm} & \textbf{Output} & \textbf{Semantic}\tnote{a} & \textbf{NRT}\tnote{b} & \textbf{QoS}\tnote{c} \\
\midrule
METIS \cite{metis}              & Post-hoc  & Tree           & \xmark & \cmark & \xmark \\
EXPLORA \cite{explora}          & Post-hoc  & Graph          & \cmark & \cmark & \xmark \\
AICHRONO. \cite{aichronolens}   & Post-hoc  & Attribution    & \cmark & \xmark & \xmark \\
ComTree \cite{comtree}          & Post-hoc  & Tree           & \cmark & \cmark & \xmark \\
SymbXRL \cite{symbxrl}          & Post-hoc  & Graph          & \cmark & \cmark & \xmark \\
SIA \cite{sia}                  & Post-hoc  & Graph          & \cmark & \cmark & \xmark \\
inRAN \cite{inran}              & Intrinsic & Formula        & \xmark & \xmark & \cmark \\
\textbf{\pname (Ours)} & \textbf{Intrinsic} & \textbf{Formula\&Rule} & \cmark & \cmark & \cmark \\
\bottomrule
\end{tabular}
\begin{tablenotes}
\footnotesize
    \item[a] Semantic high-level representations (\cmark) rather than raw KPMs (\xmark).
    \item[b] Enable Near-Real-Time (Near-RT) execution (\cmark) otherwise (\xmark).
    \item[c] Provides QoS guarantees (\cmark) otherwise (\xmark).
\end{tablenotes}
\end{threeparttable}
\end{table}

\textbf{XAI for Networking.}
Research on network interpretability has evolved from basic structural conversion toward real-time symbolic reasoning. METIS \cite{metis} introduces the distillation of deep neural network (DNN) policies into decision trees and hypergraphs to produce human-readable rules. However, these static conversions often lack the wireless context required for root-cause analysis. EXPLORA \cite{explora} addresses this gap by using attributed graphs to link agent actions to the input state space. To handle the temporal complexity of mobile traffic, AICHRONOLENS \cite{aichronolens} integrates classical XAI with time-series analysis. SIA \cite{sia} further introduces an influence score to audit forecast-augmented agents within sub-millisecond windows, and reveals issues such as temporal misalignment. SYMBXRL \cite{symbxrl} and ComTree \cite{comtree} explore FOL and large language model (LLM)-guided metrics to improve the readability of explanations. However, these methods remain reactive to distribution shifts. inRAN \cite{inran} addresses the non-stationarity issue by employing Kolmogorov-Arnold Networks (KANs) as interpretable surrogate models for online Bayesian learning at the cost of moderate computational overhead. 
Table~\ref{tab:xai_comparison} compared \pname against existing XAI methods.

\textbf{Knowledge Distillation for DRL.}
Knowledge distillation transfers behavior from a high-capacity teacher model into a compact student model. Policy distillation was first introduced for deep Q-networks \cite{rusu2015policy}. Subsequent symbolic extensions, such as deep symbolic policies (DSP) \cite{dsp} and NUDGE \cite{nudge}, further distill neural controllers into closed-form expressions and differentiable first-order logic (FOL) rules, respectively. 
Unlike prior work on classical control or game benchmarks, \pname brings symbolic policy distillation to the O-RAN control plane by grounding concepts in 3GPP/O-RAN specifications and enforcing QoS and Near-RT constraints through symbolic action shielding for safe deployment.

\section{Conclusion}
\label{Conclusion_7}
In this paper, we presented \pname, a system for transforming opaque DRL-based control in O-RAN into interpretable, verifiable, and deployable policies through symbolic distillation. By introducing a spec-grounded conceptizer, \pname bridges low-level KPMs and high-level operator-defined concepts, enabling human-aligned representations of network state. Building on this abstraction, the symbolic distiller converts neural policies into compact mathematical expressions and logical rules, while preserving fidelity to the original DRL behavior. To ensure safe deployment, the symbolic action shielding module enforces system-level constraints through projection and retrieval mechanisms.
Overall, this work highlights a practical path toward trustworthy AI-driven network control by combining learning-based optimization with symbolic reasoning and domain knowledge.

\bibliographystyle{plain}
\bibliography{ref}

\clearpage
\appendix

\end{document}